\documentclass[aps,pra,amssymb,twocolumn,showpacs,supersriptaddress,groupeaddress]{revtex4-1}
\usepackage{amssymb}
\usepackage{amsmath}
\usepackage{graphicx}
\usepackage{float}
\usepackage{dcolumn}
\usepackage{bm}
\usepackage{natbib,hyperref}
\usepackage{hyperref}

\usepackage{color}
\hypersetup{colorlinks=true, % false: boxed links; true: colored links
    linkcolor=red,          % color of internal links (change box color with linkbordercolor)
    citecolor=magenta,        % color of links to bibliography or green
    filecolor=gree,      % color of file links or magenta
    urlcolor=blue           % color of external links or cyan
}

\begin{document}

%\preprint{APS/123-QED}

\title{ Efficient steady state entanglement generation in strongly driven coupled qubits}

\author{Ana Laura Gramajo$^{1,3}$, Daniel Dom\'{\i}nguez$^{1}$ and Mar\'{\i}a Jos\'{e} S\'{a}nchez $^{1,2}$}

\affiliation{$^{1}$ Centro At{\'{o}}mico Bariloche and Instituto Balseiro (Universidad Nacional de Cuyo), 8400 San Carlos de Bariloche, Argentina.\\
$^{2}$ Instituto de Nanociencia y Nanotecnología (INN), CONICET-CNEA, Argentina. \\
$^{3}$ The Abdus Salam International Center for Theoretical Physics, Strada Costiera 11, 34151 Trieste, Italy
}

\date{\today}%
\begin{abstract}
	
We report on a  mechanism to optimize the generation of steady-state entanglement in a system of coupled qubits driven by microwave fields.
Due to the interplay between Landau-Zener-St\"uckerlberg pumping involving three levels and a subsequent fast relaxation channel,
 which is activated by tuning the  qubits-reservoir couplings, a  maximally entangled state can be populated. This  mechanism does not require from the fine-tuning of multiphoton-resonances but depends on the sign of the qubit-qubit coupling. In particular, we find that by a proper design of the system parameters and the driving protocol, the two-qubits steady-state concurrence can attain  values close to 1 in a wide range of driving amplitudes. Our results may be useful to gain further insight into entanglement control and manipulation in  dissipative  quantum systems exposed to strong driving.

\end{abstract}
%\pacs{74.50.+r,82.25.Cp,03.67.Lx, 03.65.Ud,42.50.Hz}
\maketitle

\section{Introduction}

The creation of on demand  entangled states for coupled qubits systems exposed to   dissipative environments  is a challenge requirement
to be fulfilled for most  quantum operations. It is thus crucial to study the generation and control of entanglement  in  open quantum systems. Several proposal have  shown that  noise and  coupling to the environment can be used in certain situations 
to obtain  steady-state entanglement  from dissipative processes.
Besides the specific design of the qubit-qubit interaction, strategies   based on engineering  the quantum reservoir or the system-reservoir coupling in 
 order to stabilize entanglement and to achieve quantum controlled state preparation  have been tested \cite{kraus_2008,verstraete_2009, barreiro_2011,krauter_2011,lin_2013,reiter_2013,shankar_2013,leghtas_2013,kienzler_2015,kimchi_2016,tacchino_2018}.
These approaches require an external coherent driving
field, and  following   this route  two different regimes have  been explored so far. 

For weak  resonant  driving, experimental demonstrations of entanglement stabilization are  based on tailoring the relaxation rates in order 
to generate a nontrivial  non-equilibrium dynamics which  leads to a highly entangled steady state. 
Examples of these strategies have been followed in atomic
ensembles \cite{krauter_2011},  trapped ions \cite{barreiro_2011,lin_2013,kienzler_2015} and superconducting qubits \cite{shankar_2013,leghtas_2013,kimchi_2016,quintana_2013,champagne_2018} and on general basis involve three levels
and the tuning of specific resonances among them.
More recently,  Ref. \cite{li_2020}  proposed  a frequency-modulation of a periodically 
pump laser to  achieve an accelerated formation
of dissipative entangled steady state in Rydberg atoms.

For non resonant and {\it large} amplitude  periodic drivings, a mechanism relying on the 
 amplitude-modulation of the  periodic (ac) signal was recently proposed for generating dissipative steady-state entanglement in a solid-state
 qubits system  interacting with  a thermal bath \cite{gramajo_2018}.  In analogy to well known protocols used to study 
  Landau-Zener-St\"uckelberg (LZS) interferometry, multiphoton resonances  \cite{oliver_2005,sillanpaa_2006,berns_2006,rudner_2008,izmalkov_2008,shevchenko_2010,wilson_2010,dupont_2013,forster_2014,neilinger_2016} and bath-mediated population inversion \cite{stace_2005,stace_2013,ferron_2012,ferron_2016} in two-level systems,
entanglement in the steady state has been induced  and tuned by changing the amplitude of  the ac field in a system composed of two driven  and coupled superconducting qubits. 

Interestingly, and depending on the relevant time scales,  three different scenarios for entanglement evolution have been found in Ref.\cite{gramajo_2018}:
(i) A dynamic generation of entanglement at multiphoton resonances for time scales below the decoherence time, in accordance with previous  results for
non dissipative evolutions \cite{sauer_2012,gramajo_2017}, (ii) 
{\it entanglement blackout}, or  entanglement  destruction due to decoherence with the environment, for times scales longer than decoherence  but
shorter than the relaxation time  and (iii)  the generation of steady state entanglement out but close to  specific multiphoton resonances for long times (above the relaxation time), with the possibility to enhance entanglement by tuning the driving amplitude.
As has been discussed in detail in Ref.\cite{gramajo_2018}, the generation of steady state  entanglement  requires two levels and some  fine tuning of parameters in order to be  close but out of specific resonances, a fact that could be considered as a possible limitation
for  the proposed scheme. 

The high-tunability of superconducting  qubits,  besides demonstrating the full  control of the inductive,
 capacitive and Ising-like type  of  coupling  between  qubits \cite {majer_2005, amin_2005,grajcar_2006, vanderploeg_2007}, enables to efficiently  modify the coupling strengths between each qubit and the electromagnetic environment \cite{xiang_2013,yan_2016, forn_2017,lu_2017,srinivasan_2011,hutchings_2017}.
 As we show  in this work this last tool opens a new avenue for  steady state entanglement stabilization. 
By considering that each  qubit is coupled  with  a different  strength to the  thermal bath, it may  be possible to create steady state  maximal entanglement in an efficient way without fine tuning of  a particular multiphoton resonance.
In this case, the   entanglement creation involves  three levels  and the relaxation is dominated  by a decay channel   whose contribution is negligible  in the case of identical qubit-bath couplings strengths.  

The paper is organized as follows:  in Sec.\ref{s2} we introduce the physical model and the Hamiltonian for  two-coupled qubits driven by strong ac fields. In addition we define the system-bath configuration  employed to compute the dissipative open-system dynamics. In Sec.\ref{s3} we analyze the off-resonance three-level mechanism (O3L) for entanglement creation. Other scenarios which involve the tuning of specific  resonance conditions for entanglement generation are discussed  in Sec. \ref{s4}, where   we also explain why in these cases the steady state entanglement is lower than  for the O3L. Conclusions and perspectives are given in Sec.\ref{s5}.

\section{Physical Model}
\label{s2}

We consider two coupled qubits   with Hamiltonian  ${H}_s(t)={H}_{0} + {V}(t)$, where
 \begin{equation}
 {H}_{0} = \sum^{2}_{i=1}\left(-\frac{\epsilon_{0}}{2}\sigma_{z}^{(i)} - \frac{\Delta_{i}}{2}\sigma_{x}^{(i)}\right) -\frac{J}{2}\left(\sigma^{(1)}_{+}\sigma^{(2)}_{-} + \sigma^{(1)}_{-}\sigma^{(2)}_{+}\right) 
 \nonumber
 \label{h0}
 \end{equation} and  $\sigma^{(i)}_{z,x,+,-}$ are the Pauli matrices in the Hilbert space of qubit $i$.
The parameters $\Delta_1,\Delta_2,J$ are  fixed by device design and $\epsilon_{0}$ can be controlled experimentally.
This type of Hamiltonian can be realized, for instance, in superconducting qubits \cite{berkley_2003,izmalkov_2004,majer_2005,liu_2006,zhang_2009,weber_2017,yan_2018}
where the  qubit-qubit interaction term gives rise to non trivial entangled (eigen) states of $H_0$. 
The additional term 
$${V}(t)=-A\cos(\omega t)\left(\sigma^{(1)}_{z} + \sigma^{(2)}_{z} \right)/2\;,$$ 
which contains the external ac field of amplitude $A$  and  frequency $\omega$ \cite{shevchenko_2008,ilichev_2010,satanin_2012,temchenko_2011,sauer_2012, gramajo_2017} is  usually implemented to study Landau-Zener-St\"uckelberg interferometry  in   driven  qubits \cite{oliver_2005,oliver_2009}.

\begin{figure}[!htb]
	\includegraphics[width=7cm]{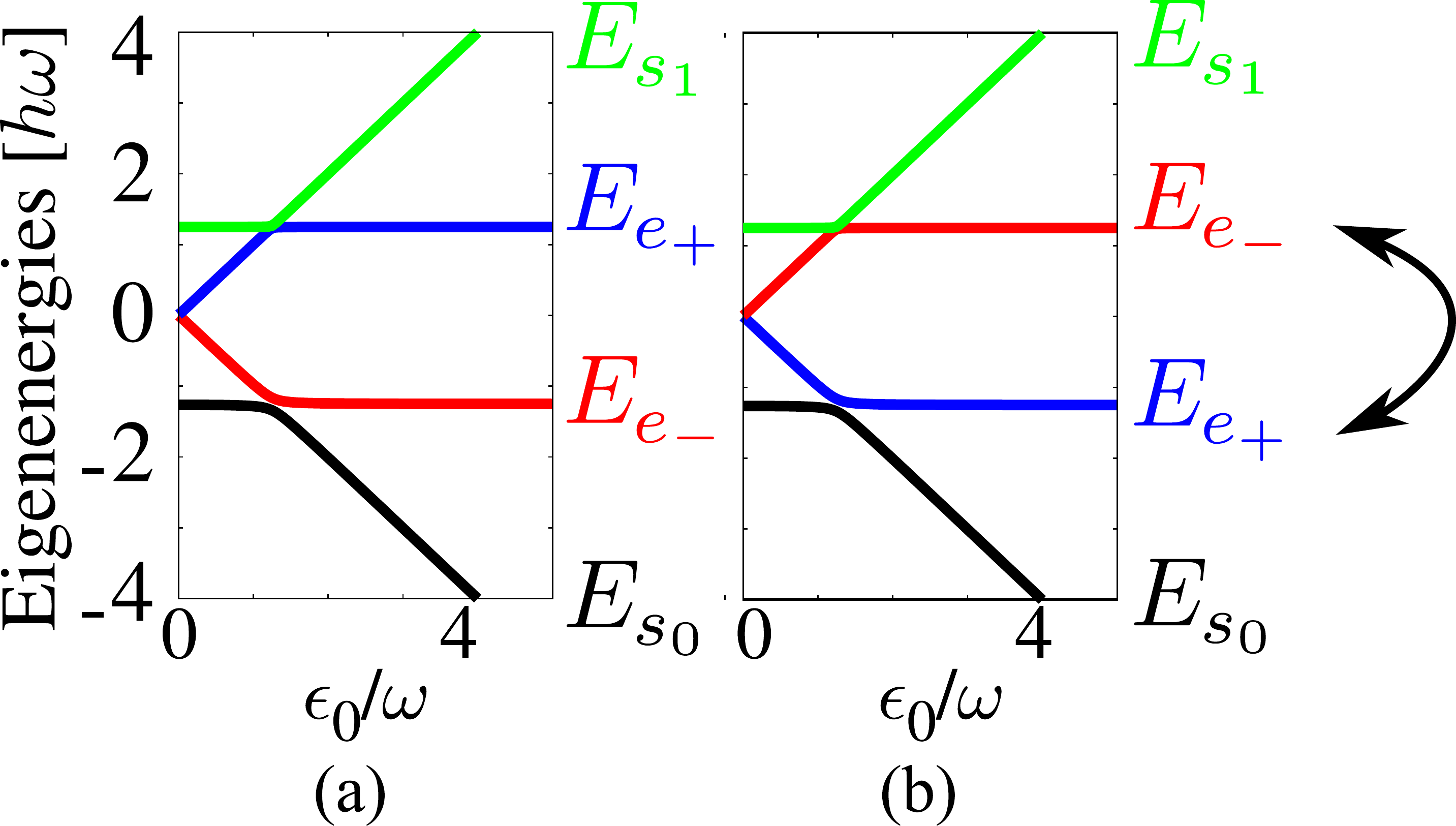}
	\caption{Eigenenergies of $H_0$  as a  function of $\epsilon_{0}/\omega$ for $J/\omega =-2.5$ (a)  and $J/\omega =2.5$ (b). Both spectra are computed for $\Delta_{2}/\Delta_1=1.5$, and $\Delta_1  /\omega= 0.1 $. For further analysis  we normalized  parameters in terms of $\omega$. See text for details.}
\label{fig:1}
\end{figure}

Figure \ref{fig:1} shows the energy spectra of $H_0$  as a function of the detuning $\epsilon_0$ for
$J <0$ [Fig.\ref{fig:1}(a)] and $J >0$ [Fig.\ref{fig:1}(b)].
In both cases, the  Hamiltonian $H_0$ for $\Delta_1,\Delta_2\ll\epsilon_0$, has two entangled eigenstates $|e_{\pm}\rangle\approx \frac{1}{\sqrt{2}}(|01\rangle \pm|10\rangle)$ (in the basis spanned by the eigenstates of  $\sigma^{(1)}_{z}\otimes\sigma^{(2)}_{z}$) with eigenenergies $E_{e\pm}\approx\mp |J|/2$, and two separable  eigenstates $|s_0\rangle\approx|00\rangle$ and  $|s_1\rangle\approx|11\rangle$, with eigenenergies $E_{s_0}\approx-\epsilon_0$ and $E_{s_1}\approx\epsilon_0$, respectively (see  Appendix \ref{sec:B} for  explicit expressions of  the eigenstates of $H_{0}$ computed using perturbation theory). 

In the following we label the states  by $|k\rangle$, with $k=0,..3$, according to their energy ordering. For instance,  the ground state is  $|0\rangle$, the first excited state is  $|1\rangle$, and so on. The energy ordering of the separable and entangled states depends on the sign of $J$ and on the value  of $\epsilon_0$ relative to  $\epsilon_c=|J|/2$, as can be seen in Fig. \ref{fig:1}. Notice that the ground state  is entangled ($|0\rangle\approx|e_{\mp}\rangle$)  for $|\epsilon_0|<\epsilon_c$ and separable ($|0\rangle\approx|s_0\rangle$) for $|\epsilon_0|>\epsilon_c$.  

To analyze  the open system dynamics it is customary to model the thermal environment  by an harmonic-oscillator bath described by a Hamiltonian ${H}_{b}$ 
and the coupling between the system and the bath by  ${H}_{sb}$,  being the global Hamiltonian   ${\cal H}(t)={H}_{s}(t) +{H}_{b} + {H}_{sb}$.
We choose  ${H}_{sb}= \mathcal{A} \otimes\mathcal{B}$, being $\mathcal{B}$ an observable of the bath and
  \begin{equation}
  \mathcal{A}= \gamma_{1}\sigma^{(1)}_{z} + \gamma_{2}\sigma^{(2)}_{z}, 
  \label{eq:sys_obs}
 \end{equation} 
the system operator which, under the assumption of  weak system-bath interaction, is taken  linear in the coupling strengths $\gamma_{1,2}$. 
Notice that other functional forms for the $\mathcal{A}$ operator can be also  considered, but we use this one in order to model a realistic situation for SC-qubits coupled to the electromagnetic environment, that  we consider as a thermal bath at  temperature $T_b$ and with an Ohmic spectral density $J(\Omega)=\kappa\Omega e^{-|\Omega|/\omega_{c}}$. 

The dynamics of the reduced density matrix  of the two coupled qubits $\rho(t)={\rm Tr}_{b}\left(\rho_{tot}\right)$ is  obtained by tracing out  the degrees of freedom of the  bath from the global density matrix $\rho_{tot}$.  
We  numerically  solve the corresponding quantum master equation for the reduced density matrix,  under the Floquet-Born-Markov approach \cite{kohler_1997,blattmann_2015,kohler_2018,ferron_2016,gramajo_2018}, which allows the treatment of  open systems under periodic drivings of arbitrary strength and frequency (more details are given in App.(\ref{sec:A})). The  entanglement between the two qubits  is quantified by the concurrence, which is defined  as $C=\text{max}\{ 0, \lambda_{4}-\lambda_{3}-\lambda_{2}-\lambda_{1} \}$, being  $\lambda_{i}$'s  the real  eingenvalues  in decreasing order of the matrix $R=\sqrt{\sqrt{\rho}\tilde{\rho}\sqrt{\rho}}$, with $\tilde{\rho}=\sigma^{(1)}_{y}\otimes\sigma^{(2)}_{y} \rho^{*}\sigma^{(1)}_{y}\otimes\sigma^{(2)}_{y}$ \cite{wootters_1998}. From  $\rho(t)$ and $\overline{\rho_{\infty}}$ ($\overline {<..>} $ means averaged over one driving period $ 2 \pi/\omega$) we compute the time dependent and  the steady state concurrences, $C(t)$ and  $C_\infty$ respectively, when  the system is initially prepared in   a separable  (ground) state of $H_0$.

\section{Steady state entanglement generation: Off-resonance three levels mechanism}
\label{s3}

In a previous work \cite{gramajo_2018} we have shown that for a strong ac driving, i.e  for  large enough  amplitudes $A$, steady state
entanglement can be generated near some multiphoton resonances, when the  initial  ground state is disentangled (a condition that in our model corresponds to  $|\epsilon_0|>\epsilon_c$).

The  entanglement generation studied in  Ref.\cite{gramajo_2018}  assumed  that both  driven qubits were coupled  to the  thermal bath with the same strength $(\gamma_1= \gamma_2$  in Eq.(\ref{eq:sys_obs})).

In the following, we extend the analysis by considering   $\gamma_2 = \xi \gamma_1$, being $\xi \le 1$ the parameter that quantifies the relative degree of coupling between each qubit and the thermal bath. As we will  show, for $\xi \neq 1$   and  $\Delta_1\neq \Delta_2$  an extra relaxation channel opens, providing  a new avenue  to  maximize the steady state concurrence and  enlarge significantly the region in  parameters space where entanglement can be generated. 

Hereafter we  fix  $\Delta_{2}/\Delta_{1}=1.5$,    $\Delta_{1}=0.1 \omega $, the bath  temperature  $T_{b}/ \omega=0.00467$  ($\sim$ 20 mK for typical superconducting qubits) and  $\kappa=0.001$. We also set $\gamma_{1}=1$ and $\gamma_{2}=\xi$ for numerical calculations.
\begin{figure}[!htb]
	\includegraphics[width=5.5cm]{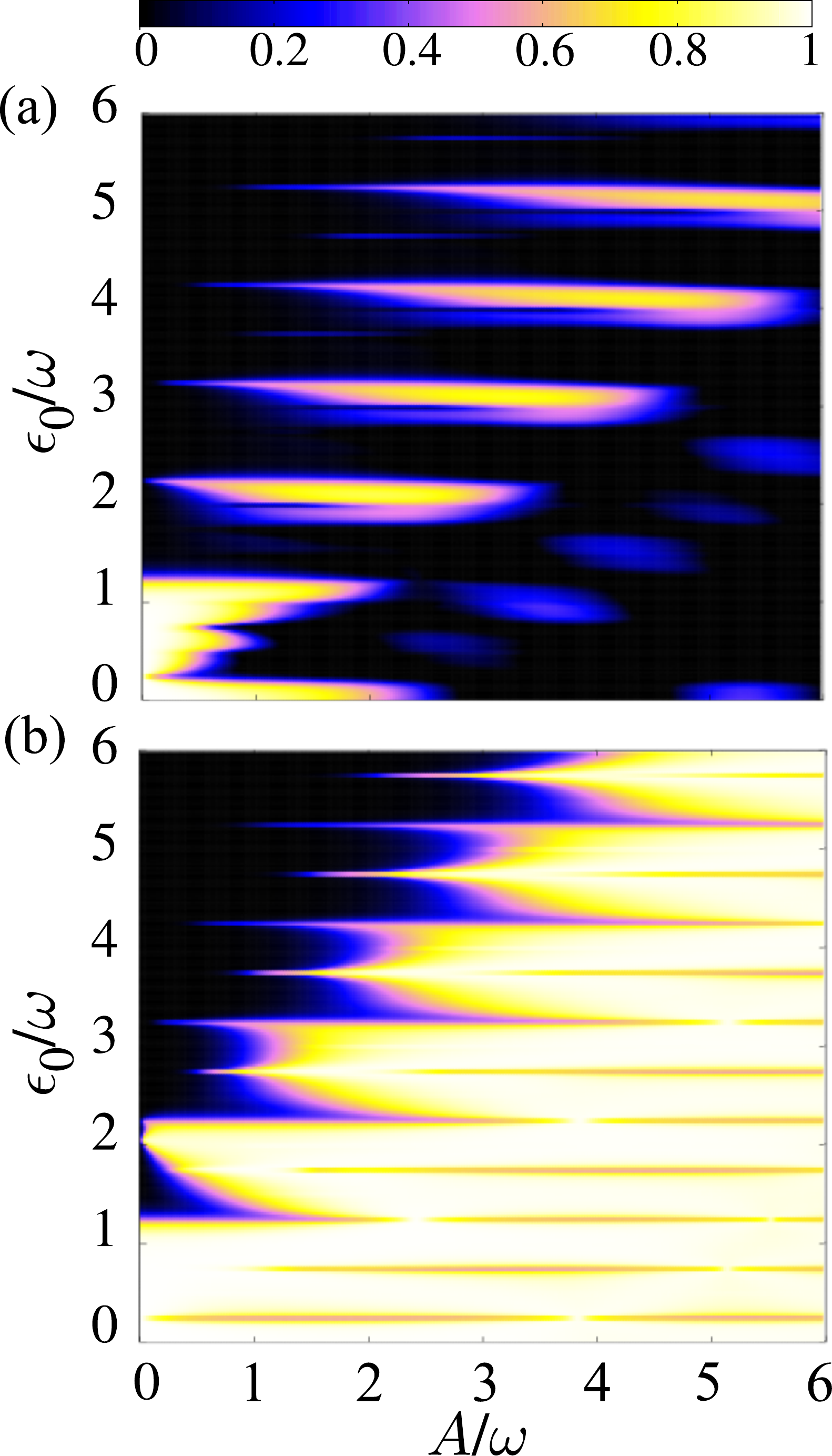}
	\caption{Colour map of  $C_{\infty}$ versus $A/\omega$ and $\epsilon_{0}/\omega$ for $\xi =1$ (a)  and $\xi =0.1$ (b), with $\gamma_{1}= \gamma$ and  $\gamma_{2}=\xi \gamma_{1}$. Both cases are computed for $J/\omega=-2.5$.
	See text for details.}
\label{fig:2}
\end{figure}

We will analyze in what follows the  results for   $J <0$ (see App.\ref{sec:D} for the case $J>0$).
Fig.\ref{fig:2} shows the intensity plot of the steady state concurrence $C_{\infty}$ as a function of $A/\omega$ and $\epsilon_{0}/\omega$ for  $J/\omega=-2.5$.  
For this case, we have chosen two different couplings in the  system operator $\mathcal{A}$ (Eq.\eqref{eq:sys_obs}):  $\xi=1$ (Fig.\ref{fig:2} (a))-already studied in Ref.\cite{gramajo_2018}, and $\xi=0.1$ (Fig.\ref{fig:2} (b)) corresponding to very  dissimilar  qubit-bath couplings.  As it is evident, the steady-state concurrence exhibits striking differences.

For $\xi=1$, the structure of $C_{\infty}$  shown in Fig.\ref{fig:2}(a) has been explained in \cite{gramajo_2018}.   There is entanglement generation for $|\epsilon_0|>\epsilon_c$, as a result of a near-resonance mechanism involving two levels (N2L) \cite{ferron_2012,ferron_2016}, mediated by the interplay of the external driving and the relaxation process that induce the required  population inversion.  In particular, the N2L  takes place  for $\epsilon_0 > \epsilon_c$
near to  (but out off) a specific multiphoton resonance condition, that sets the energy difference between the separable ground state $|0\rangle \approx |s_{0}\rangle$   and the entangled state $|e_{-}\rangle$  to $\Delta E_{se} \sim\epsilon_{0} - |J|/2= n\omega$ with $n\in\mathbb{Z}$ \cite{gramajo_2018}.
For these cases, and  by adequately
tuning a range of  driving amplitudes $A$, the system can be  excited from the initial  ground state to a virtual multiphoton state,  which in the steady state relaxes to the maximally entangled    Bell's state $|e_{-}\rangle$ (notice  that this implies having attained population inversion).
Thus  a steady concurrence $C_{\infty}\lesssim1$ is obtained near these resonances by tuning the amplitude $A$, as it is displayed in  Fig.\ref{fig:2}(a).

For  $\xi=0.1$ the steady state concurrence exhibits a very different behavior:   $C_{\infty} \simeq 1$ in Fig.\ref{fig:2}(b)  over almost all the available parameter space $\{A,\epsilon_{0}\}$, without  requiring from a specific ``close to a resonance" condition. Notice that  the region   $|\epsilon_{0}| < \epsilon_c$  is entangled  for amplitudes  $A\rightarrow 0$, as the initial ground state is entangled  for  detunings satisfying the above condition.  

As we will discuss below, this  new behavior results from an {\it off-resonance three level}  (O3L) mechanism based on: (i) Landau-Zener pumping  from the ground state $|0\rangle$  to an ancillary excited state $|2\rangle$ and (ii) fast relaxation from  the ancillary  $|2\rangle$ to the entangled state $|1\rangle \approx |e_{-}\rangle$. The sequence of transitions  $|0\rangle \rightarrow |2\rangle \rightarrow |1\rangle $ leads to a continuous transfer of population from the ground state  $|0\rangle\approx|s_{0}\rangle$ to the entangled state $|1\rangle\approx|e_{-}\rangle$, thus giving a steady state with concurrence $C_{\infty}\simeq 1$.

To illustrate the O3L mechanism $|0\rangle\rightarrow|2\rangle\rightarrow|1\rangle$, we  start by computing  for $\xi=0.1$ the diagonal elements (populations) of the two qubits reduced density matrix $\rho_{kk}(t)$ ($k=0,..3$)  as a function of time $t$, in a typical off-resonance case $\epsilon_{0}/\omega=3.7$, and for a driving amplitude $ A/\omega=3.8$.
 \begin{figure}[!htb]
\includegraphics[width=8.5cm]{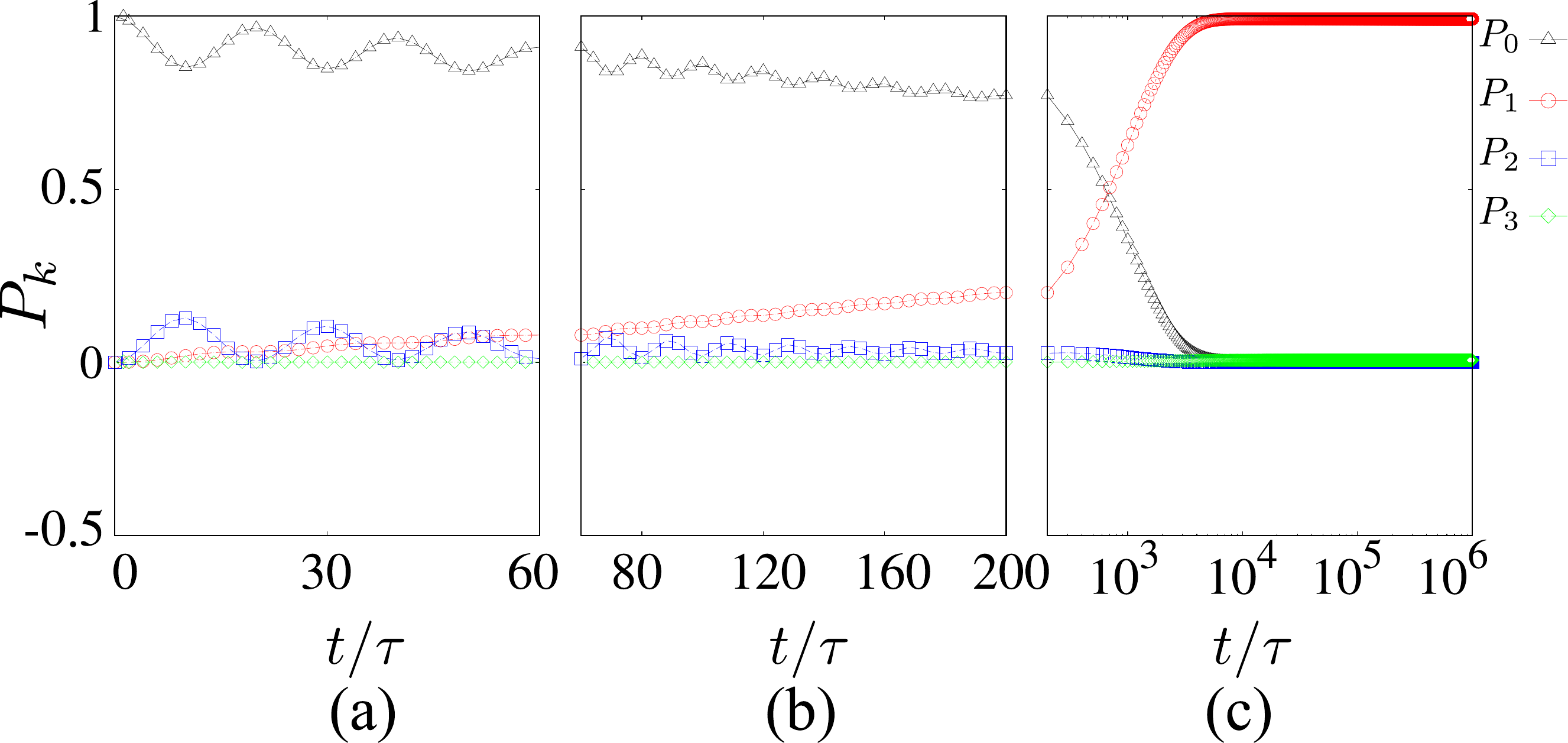}
\caption{Plots of the populations $P_{k}$, computed in the eigenstates basis of $H_0$,  as a function of normalized time $t/\tau$  ($\tau=2\pi/\omega$) for  $\xi=0.1$.  Plots (a) and (b) are in  linear  time scale, while plot (c)  is in logarithm scale.  The detuning is $\epsilon_{0}/\omega=3.7$ (off-resonance) and the driving amplitude $A/\omega=3.8$. Other parameters are the same as in Fig.\ref{fig:2}.} 
\label{fig:3}
\end{figure}

 As it is shown in Fig.\ref{fig:3}(a), the  short-time dynamics  induced by the driving mainly involves the coherent evolution  of two states: the initial ground state $|0\rangle $ and the second excited state $|2\rangle$, to which  population is transferred  via the driving induced Landau-Zener transitions at the energy levels avoided crossing ({\it Landau-Zener pumping}). 
 A necessary  condition to accomplish this is that the driving amplitude A  must be enough to reach the avoided crossing at $\epsilon_0 = \epsilon_c$, i.e.    $ A >  A_c \equiv |\epsilon_0| - \epsilon_c$. Therefore,  $A\sim A_c$ is  the characteristic crossover  amplitude   necessary to 
 activate the Landau-Zener pumping  here described ($|0\rangle \rightarrow |2\rangle$ transfer).
 
 For the present case of Fig.\ref{fig:3}(a)  is  $A=3.8\omega > A_c= 2.45 \omega$  and thus Landau-Zener pumping is active. Notice that since $\epsilon_{0}/\omega=3.7$ does not correspond to a  resonance  among these two states,  there is only a partial transfer of population from $|0\rangle \rightarrow |2\rangle$.

 As time increases  [Fig.\ref{fig:3}(b)],  a direct transition from the ancillary state $|2\rangle$ to  the state $|1\rangle$  takes place. The population $\rho_{11}(t)$ of the first excited  state starts to grow   while the  decay of the $\rho_{22}(t)$  and $\rho_{00}(t)$ populations is evident. Finally, for long times  after full relaxation, the entangled first excited state $|1\rangle\approx|e_{-}\rangle$ is fully populated, see Fig.\ref{fig:3}(c).  This {\it fast relaxation} transition is possible  
whenever the relaxation mechanism is dominated by the decay rate $\Gamma_{12}$ connecting  the   states $|2\rangle  \rightarrow |1\rangle$, as we will analyze below.

A first and straightforward estimate  of the transition rates between the eigenstates $|l\rangle$ and $|k\rangle$  can be obtained from a Fermi Golden rule  (FGR) calculation 
\begin{equation}
\begin{aligned}
    \Gamma_{kl} = \frac{2\pi}{\hbar} g(E_{lk})|\langle l |\mathcal{A} | k \rangle|^{2} ,
\end{aligned}
\label{eq:fermi_app}
\end{equation} where $\mathcal{A}$ is the observable of the system defined in Eq.\eqref{eq:sys_obs}, $E_{lk}=E_l-E_k$, and $g(E)$ accounts for the bath spectral density and thermal factors (see App.\ref{sec:C}). Since we are considering low temperatures, thermal excitations are negligible and thus the relevant decay rates are  $\Gamma_{kl}$ for  $l > k$.
 A perturbative calculation for $\Delta_1,\Delta_2\ll|\epsilon_0|$   gives for $\epsilon_0 >\epsilon_c$ (see App.\ref{sec:C} for the complete derivation) 
\begin{equation}
\begin{aligned}
&\Gamma_{12}\propto  (1- \xi)^2,\\  
&\Gamma_{02} \propto  \left(\frac{\bar{\Delta}}{\epsilon_{c}}\right)^2(1+ \xi)^2 ,\\
& \Gamma_{23} \propto  \left(\frac{\bar{\Delta}}{\epsilon_{c}}\right)^2(1+ \xi)^2 , \\
& \Gamma_{01} \propto  \left(\frac{\bar{\Delta}}{\epsilon_c }\right)^2(1- \xi)^2
\end{aligned}
\label{eq:rates}
\end{equation} with $\bar{\Delta}= (\Delta_{1} + \Delta_{2})/ 2$, neglecting terms depending on $|\Delta_1-\Delta_2|<\bar{\Delta}$, and prefactors (of the order of unity) depending on $\epsilon_0/\epsilon_c$.

\begin{figure}[!htb]
\includegraphics[width=5.5cm]{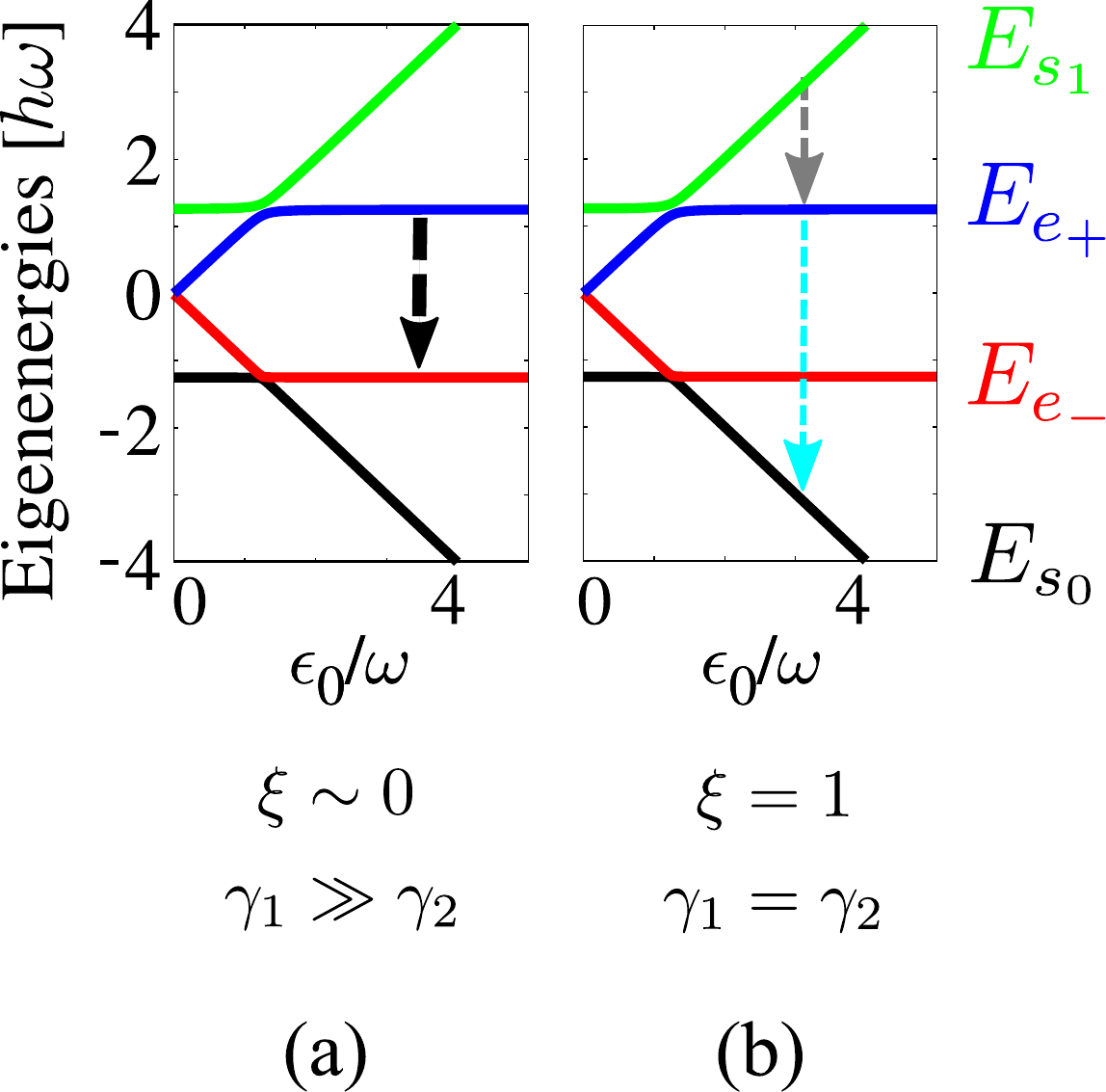}
\caption{Eigenenergy spectrum as a function of  $\epsilon_{0}/\omega$. The most relevant relaxation processes for  $\xi \sim 0$ (a) and $\xi =1$ (b) are  sketched by arrows. The results correspond to   $J/\omega=-2.5$,  $\epsilon_{0}=3.7\omega$ and  no driving, i.e. $A/\omega=0$. Other parameters are the same as in Fig.\ref{fig:2}.}
\label{fig:4}
\end{figure}

\begin{figure}[!htb]
\centering
\includegraphics[width=6cm]{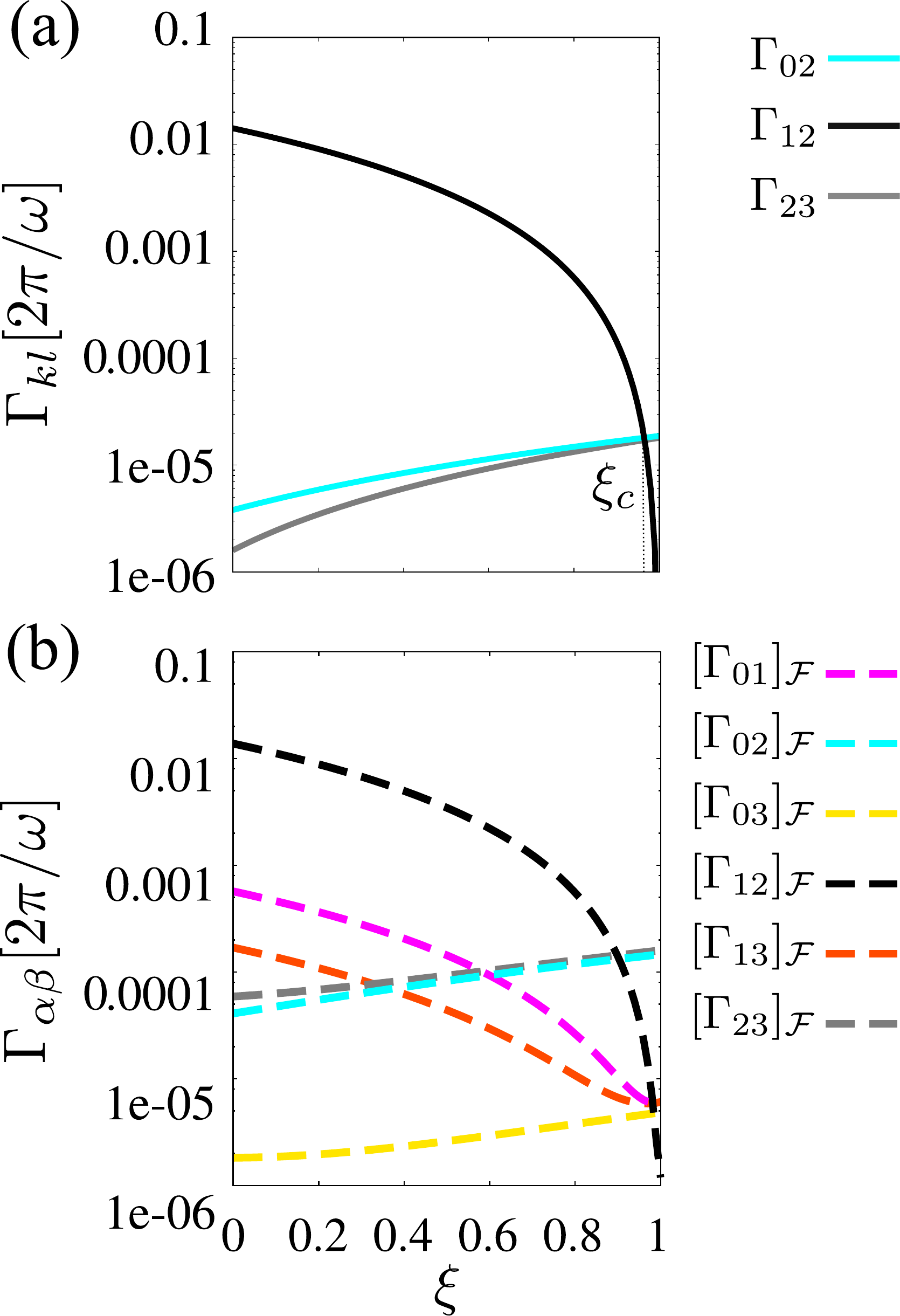}
\caption{Plots of the  transition rates as a function of $\xi$ for the off-resonance case $\epsilon_{0}=3.7\omega$. (a) Analytical results  $\Gamma_{kl}$ computed in the eigenstate basis,  employing Eq.\eqref{eq:rates_app} of App.\ref{sec:C}. The value of $\xi_{c}$ is indicated. (b) $\Gamma_{\alpha\beta}$ computed in the Floquet basis  for $A/\omega=3.8\omega$. Other parameters are the same as in Fig.\ref{fig:2}. 
}
\label{fig:5}
\end{figure}

From Eq.(\ref{eq:rates}) it is clear that for $\xi=1$, $\Gamma_{12}$  vanishes and the largest transition rates are ${\Gamma}_{23}$ and  ${\Gamma}_{02}$. In this case the system  will tend to relax to the ground state $|s_0 \rangle$,  as  Fig.\ref{fig:4}(b) shows schematically. On the other hand,  for  $\xi \rightarrow 0$ the rate   ${\Gamma}_{12}$  attains its maximum value and is by far the largest one, providing the fast relaxation mechanism that  results in the  entangled state $| e_{-} \rangle$  being fully populated, also shown schematically in  Fig.\ref{fig:4}(a). We plot in Fig.\ref{fig:5}(a) the most relevant relaxation rates as a function of $\xi$,  estimated with the FGR 
and using the full expressions given in  App.\ref{sec:C}.  As can be seen, there is a characteristic $\xi=\xi_c$  such that $\Gamma_{02},\Gamma_{23}\gg \Gamma_{12}$  for  $\xi>\xi_c$ and $\Gamma_{12} \gg  \Gamma_{02},\Gamma_{23}$,  for $\xi<\xi_c$. From  Eq.(\ref{eq:rates}), one obtains $1-\xi_c \propto \bar{\Delta}/\epsilon_c$. 

To simplify the analytical calculations we have neglected the  dependence of  $\Gamma_{kl}$ on the driving amplitude, since we have computed the transition rates among eigenstates of the undriven Hamiltonian $H_0$. However, the  natural basis to compute the transition rates in the case of a strongly driven system is the Floquet basis ${\mathcal{F}}$, in which  the density matrix  in the steady state  becomes diagonal \cite{ferron_2016}. 
Following this route, in Fig.\ref{fig:5}(b) we plot the transition  (relaxation) rates ${\Gamma}_{\alpha\beta}$  computed numerically in the Floquet basis ${\mathcal{F}}$ \cite{ferron_2012,ferron_2016,gramajo_2018}   as a function of $\xi$,  for the considered  off-resonant situation, $\epsilon_{0}/\omega=3.7$. 

To each Floquet state we can associate the $H_0$ eigenstate to which it tends for $A\rightarrow 0$.
We have then labeled the Floquet states $\alpha, \beta, \gamma, \delta$ following  the same  ordering  than  the eigenstates of $H_0$
(this is a reasonable choice since  quasienergies do not cross  for  out of resonance conditions \cite{shevchenko_2010, grifoni_2010,ferron_2012}). 
Notice that the  rates in the Floquet basis, $[{\Gamma}_{12}]_{\mathcal{F}}$, $[{\Gamma}_{02}]_{\mathcal{F}}$ and $[{\Gamma}_{23}]_{\mathcal{F}}$, have a  functional dependence on $\xi$ similar  to  the estimates given in  Eq.\eqref{eq:rates},
and shown in Fig.\ref{fig:5}(a). 
In particular for  $\xi=1$  the largest transition rates are $[{\Gamma}_{23}]_{\mathcal{F}}$ and  $[{\Gamma}_{02}]_{\mathcal{F}}$, while   for  $\xi \rightarrow 0$ the relaxation process is dominated by   $[{\Gamma}_{12}]_{\mathcal{F}}$, in correspondence  with the previous description. 
As it is  discussed in App.\ref{sec:C}, for out of resonance situations  the rates computed in the Floquet  basis, $\Gamma_{\alpha\beta}$, give essentially the same qualitative information regarding  the main relaxation processes than the rates computed in the eigenbasis of $H_0$, $\Gamma_{kl}$.

Taking into account that the most accurate description is in terms of the Floquet transition rates, the crossover $\xi_c$  should be defined from the condition $[{\Gamma}_{12}]_{\mathcal{F}} = {[\Gamma}_{02}]_{\mathcal{F}}$.  Thus in general  $\xi_c$ will depend on the driving amplitude $A$.

\begin{figure}[!htb]
\includegraphics[width=6cm]{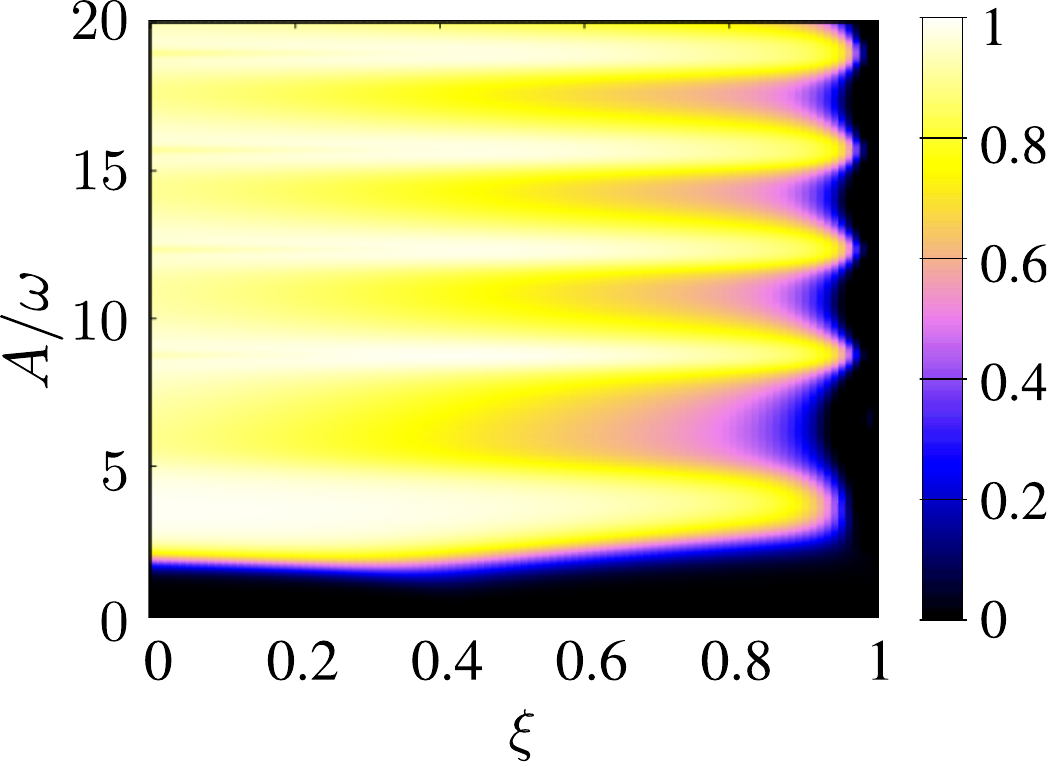}
\caption{Intensity plot of $C_{\infty}$ versus $\xi$ and $A/\omega$ for the off-resonance case $\epsilon_{0}/\omega=3.7$. Other parameters are the same as in Fig.\ref{fig:2}.}
\label{fig:6}
\end{figure}

From  the above discussion we stress that  to attain the O3L mechanism  the conditions $ A \gtrsim A_c $ (Landau-Zener pumping) and $\xi\lesssim\xi_c$ (fast relaxation) have to be fulfilled simultaneously. 
This is confirmed in  Fig.\ref{fig:6}, where   $C_{\infty}$ is plotted as a function of $\xi$ and $A/\omega$ for the mentioned off-resonance condition. 
For $\xi=1$ there is no noticeable entanglement creation for all the explored  values of the  amplitude $A$ (see also Fig.\ref{fig:2}(a) for $\epsilon_{0}/\omega=3.7$), while for  $\xi < \xi_c$, a finite concurrence  $C_{\infty}>0 $ is obtained.   It is also clear from 
Fig.\ref{fig:6} that $\xi_c$ has a  modulation with $A$, as expected from the previous discussion on the Floquet relaxation rates.

We emphasize that  driving and dissipation are the two key ingredients  to generate  entanglement- as in the case of identical qubit-bath couplings  studied in Ref.\onlinecite{gramajo_2018}. However the entanglement generation here described does not  require to tune a given resonance condition but  to tailor the system-bath interaction to allow for the needed relaxation channel. 

So far we have shown that the  generation of steady state entanglement for $\xi \neq 1$  and for off-resonant situations,  relies on the O3L mechanism described along this section.  Despite we have focused on a specific value of $J= -2.5 \omega$, the O3L mechanism rules the generation of entanglement in off-resonant situations for  $\xi \neq 1$ and general values of  $J<0$, as we  discuss in App.\ref{sec:D}. 

However, the  O3L mechanism is completely suppressed  for  $J>0$. In this case the first and second excited states are switched between each other $|e_{-}\rangle \leftrightarrow |e_{+}\rangle $ ($1 \leftrightarrow 2$).  In particular for  $\xi \rightarrow 0$, in addition to $[\Gamma_{12}]_{\mathcal{F}}$, the second relevant relaxation rate becomes  $[\Gamma_{02}]_{\mathcal{F}}$, activating  the decay process $|2\rangle \rightarrow |0\rangle$ that tends to populate the ground state at long times (see  App. \ref{sec:D} for a detailed discussion).

\section{Dynamics of entanglement generation at resonances}
\label{s4}

\begin{figure}[!htb]
\centering
\includegraphics[width=5cm]{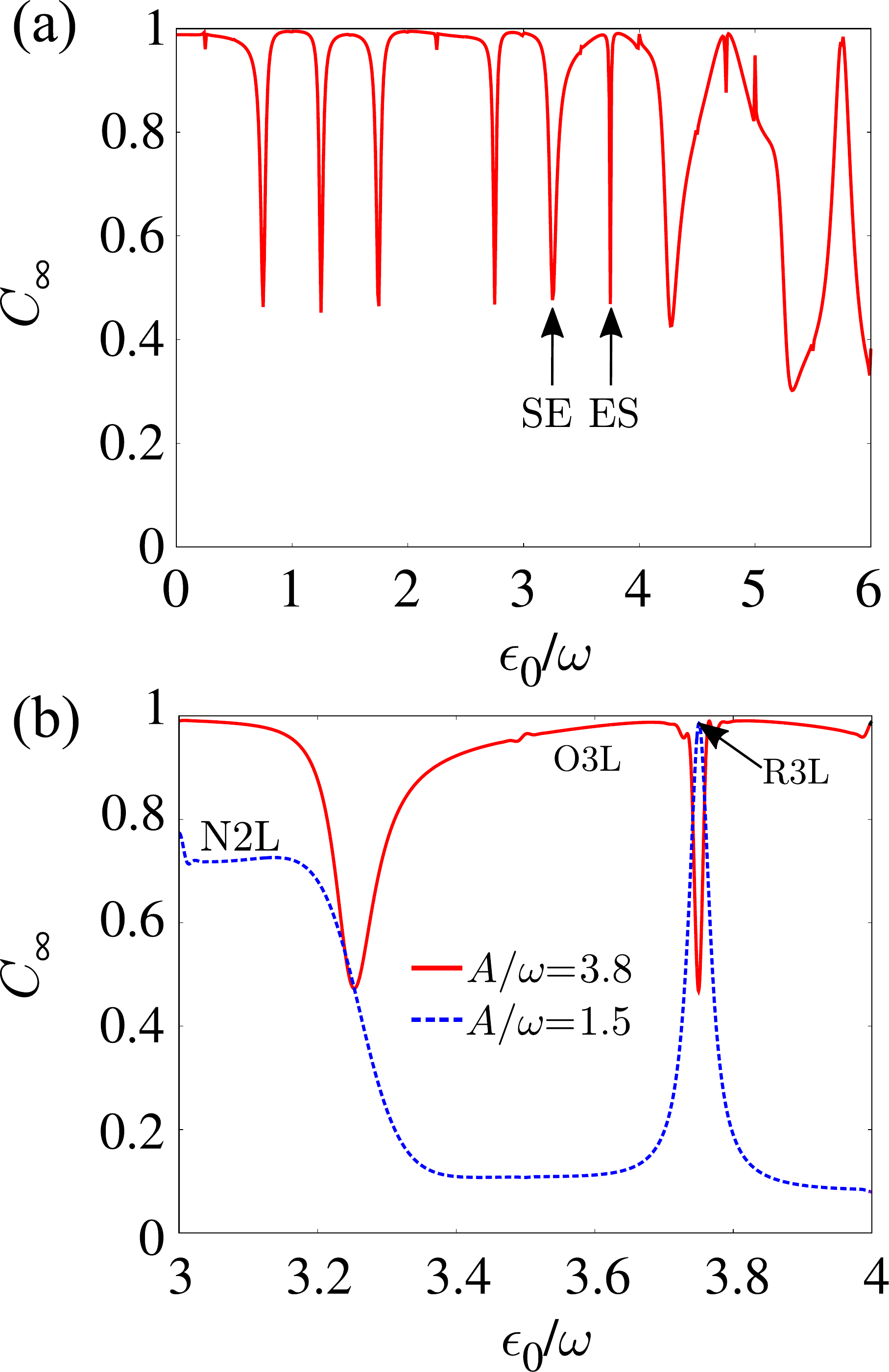}
\caption{(a) Plot of $C_{\infty}$ as a function of $\epsilon_{0}/\omega$ for the driving amplitude  $A/\omega=3.8$. The resonances   $\epsilon_{0}/\omega=3.25$ (SE) and $\epsilon_{0}/\omega=3.75$ (ES) are indicated by arrows. (b) Zoom-in plot  near the  SE and ES resonances for  amplitude  $A/\omega=3.8$ (red) and $A/\omega=1.5$ (blue).  The  labels N2L, R3L and O3L  indicate the main mechanisms for entanglement generation  described in the  text. Other parameters are the same as in Fig.\ref{fig:2}}.
\label{fig:7}
\end{figure}

The reduction in the amount of entanglement at resonances  $\epsilon_{0} \pm |J|/2 \sim m \omega$ is evident
for example in Fig.\ref{fig:2}, along the  straight lines  where   $C_{\infty} \sim 0.5$, and in  Fig.\ref{fig:7}(a), where we plot $C_{\infty}$ as a function of $\epsilon_{0}$ for $A=3.8\omega$.

We distinguish  two  types of resonances: 
(i) resonances between the separable state $|0\rangle$ and the entangled state $|1\rangle$, for $\epsilon_{0} \sim m \omega +|J|/2 $, which we name SE resonance, and
(ii) resonances between the entangled state $|1\rangle$ and the separable state $|3\rangle$, for $\epsilon_{0} \sim m \omega -|J|/2 $, which we name ES resonance.
In what follows   we  will study the detailed dynamics of entanglement generation for  two examples of these  resonances: $\epsilon_{0}/\omega=3.25$ (SE) and  $\epsilon_{0}/\omega=3.75$ (ES),  which are indicated by arrows in Fig.\ref{fig:7}(a).

\begin{figure}[!htb]
\centering
\includegraphics[width=6cm]{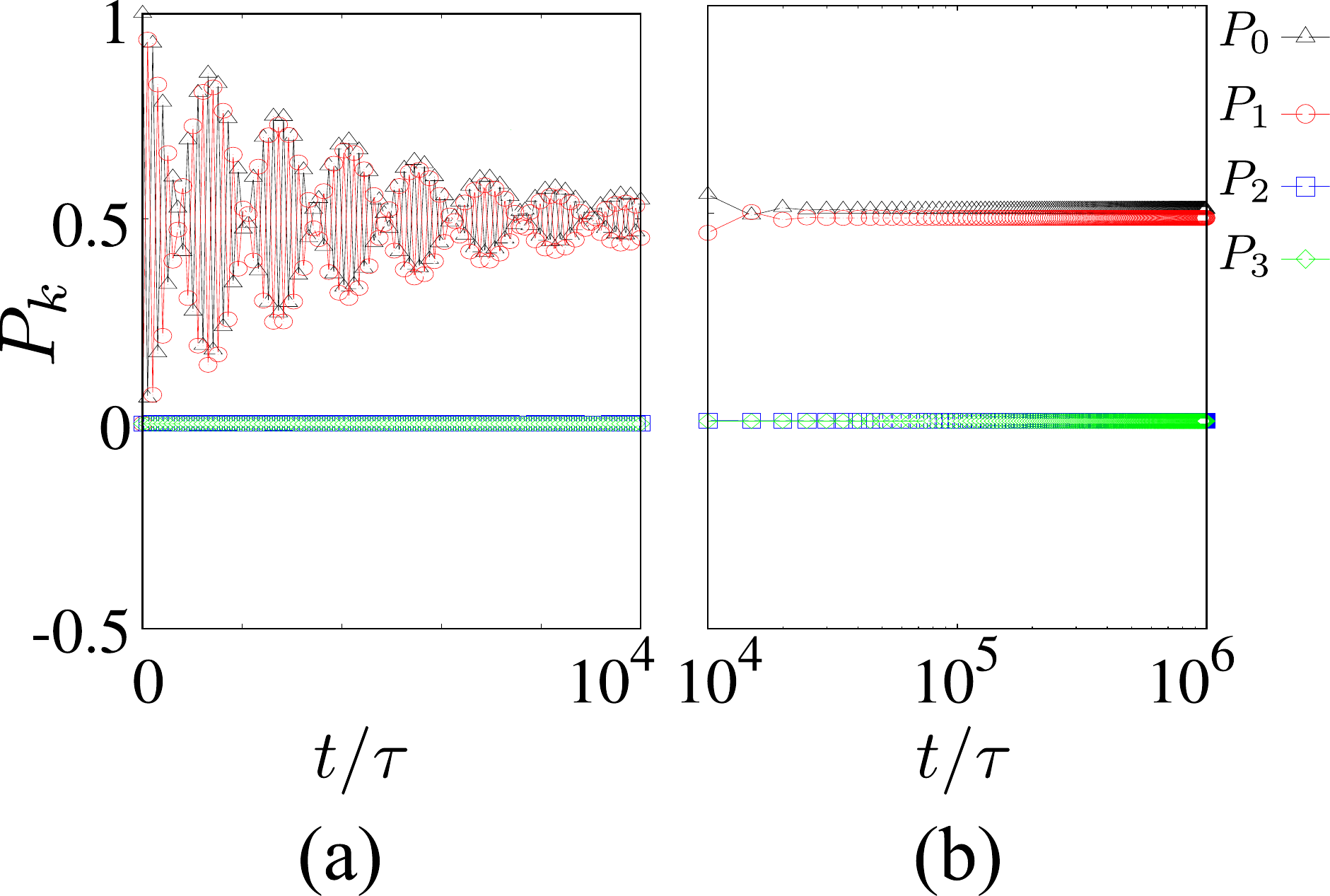}
\caption{Plots of the populations $P_{k}$, computed in the eigenstates basis of $H_0$,  as a function of normalized time $t/\tau$  ($\tau=2\pi/\omega$) for $\xi=0.1$.   The detuning is $\epsilon_{0}/\omega=3.25$ (SE resonance) and the driving amplitude is  $A/\omega=3.8$. Other parameters are the same as in Fig.\ref{fig:2}. Plot (a) is in linear  time scale and plot (b) in logarithm  time scale.}
\label{fig:12_8}
\end{figure}

The time-evolution of the populations $P_{k}(t)=\langle k|\rho(t)|k\rangle$ computed in the eigenstate basis 
 of $H_0$  is shown in Fig.\ref{fig:12_8} for $\epsilon_{0}/\omega=3.25$, which, as we mentioned, corresponds to a  multiphoton resonance condition  between the states $\{|0\rangle =|s_{0}\rangle\}$ and  $\{|1\rangle=|e_{-}\rangle \}$ (SE resonance).  Since the initial state  is the ground state ($P_0(0)=1$), at short times  there is a coherent oscillatory exchange of population with the first excited state, as shown in Fig.\ref{fig:12_8}(a). Notice that there are not intermediate populated levels ($P_{2}=P_{3}=0$), and  the coherence between $\{|s_{0}\rangle, |e_{-}\rangle \}$ dies off at longer times due to decoherence effects. The final steady state  populations are $P_{0}=P_{1}=0.5$, see Fig.\ref{fig:12_8}(b). Since  $|0\rangle$ is separable while  $|1\rangle$ is a entangled state, a partial entanglement generation is attained  with $C_{\infty} \simeq 0.5 $.

\begin{figure}[!htb]
\centering
\includegraphics[width=8.5cm]{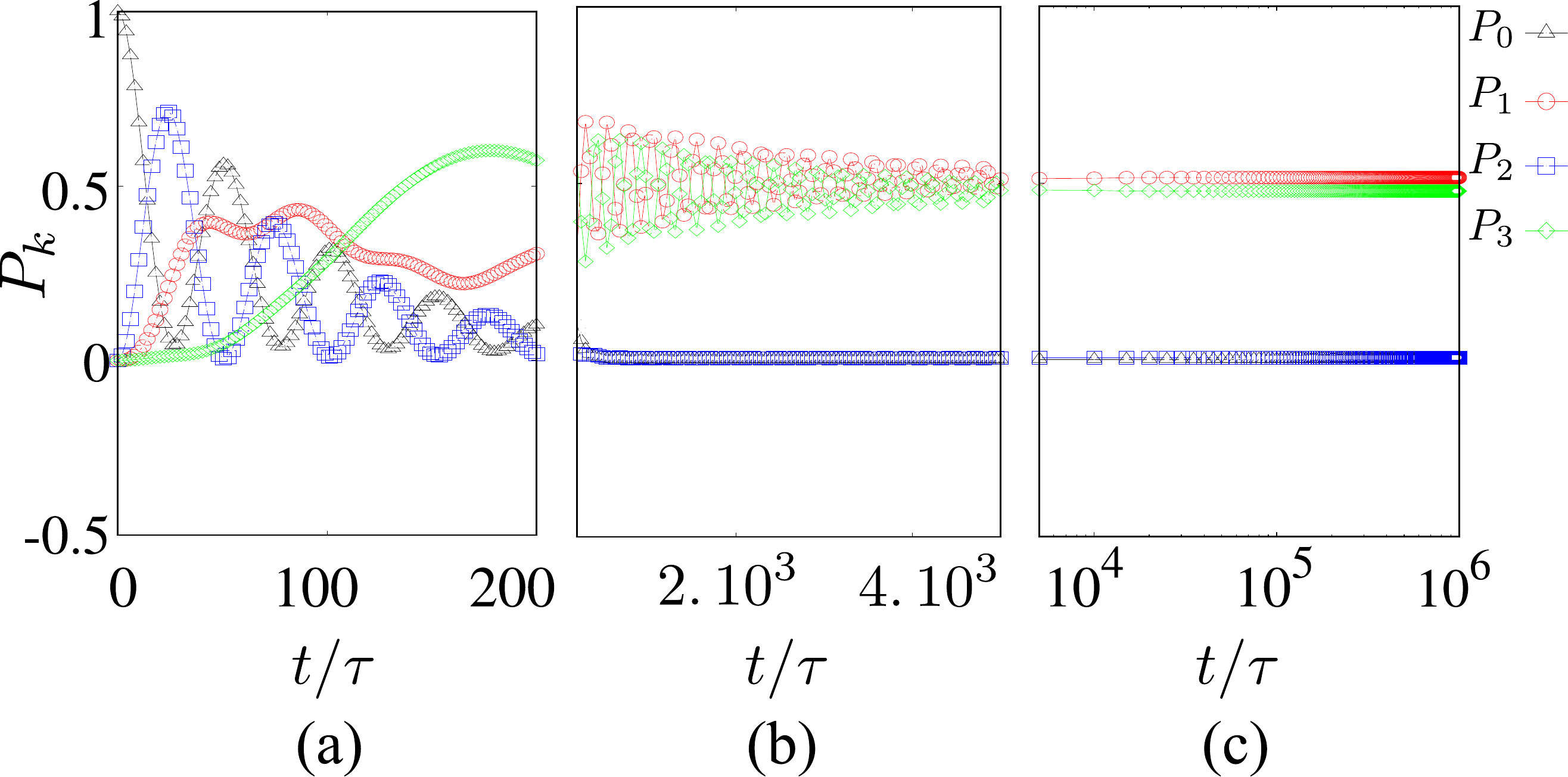}
\caption{Plots of the populations $P_{k}$ computed in the eigenstates basis of $H_0$,  as a function of normalized time $t/\tau$  ($\tau=2\pi/\omega$) for $\xi=0.1$.  The detuning is $\epsilon_{0}/\omega=3.75$ (ES resonance) and the driving amplitude is  $A/\omega=3.8$. Other parameters are the same as in Fig.\ref{fig:2}.  Plots (a) and (b) are in  linear  time scale, while plot (c)  is in logarithm scale. }
\label{fig:9}
\end{figure}

The example corresponding to the second type of  resonance (ES) is shown  for  $\epsilon_{0}/\omega=3.75$ in Fig.\ref{fig:9}, which  displays the time-evolution of the populations $P_{k}$ computed in the eigenstate basis.  In this case,
the dynamics at intermediate times is  richer because for this $\epsilon_{0}$  a simultaneous resonance condition  between the states $\{|0\rangle, |2\rangle \}$   and $\{|1\rangle, |3\rangle\}$  takes place.
For short times (as the system starts in the ground state) the populations $P_{0}$ and $P_{2}$ exhibit well-defined Rabi-like oscillations due to the resonance condition. 
As soon as the $|2\rangle$ state starts to be populated, there is a fast decay from this state to the $|1\rangle$ state, as can be observed for times $t \lesssim 100 \tau$  in Fig.\ref{fig:9}(a).
This  $|0\rangle \leftrightarrow |2\rangle \rightarrow |1\rangle$ process corresponds to the standard three level mechanism at a resonance (R3L), which induce  a net transfer of population  to the first excited state, mediated by resonant pumping from the ground state to a second excited state.
Once the $|1\rangle$ state is populated, the resonance with the $|3\rangle$ dominates the dynamics, giving place  to an oscillatory exchange between these two states for times ($t\gtrsim 200\tau$), as shown in Fig.\ref{fig:9}(b). 
These oscillations die off for $t \sim 4.10^{3}\tau$ due to  dissipative effects, and the steady-state system is ultimately reached, as  Fig.\ref{fig:9}(c) shows. The  steady-state populations are $P_{1}\sim 0.5$ and $P_{3}\sim 0.5$,  leading to a  concurrence $C_{\infty}\sim 0.5$.

The difference in the entanglement generation for the two types of resonances becomes more evident for low driving amplitudes. For $A\lesssim A_c$ the O3L mechanism is turned off since the Landau-Zener transfer to higher states is  suppressed,  and  thus the other mechanisms for entanglement generation at resonances are unveiled.  This  is well-illustrated in Fig.\ref{fig:7}(b) for the amplitude $A=1.5\omega<A_{c}$. Notice that  at one  side of the SE  resonance  ($\epsilon_{0}/\omega=3.25$) there is  entanglement generation due to the N2L mechanism involving  $|0\rangle\leftrightarrow|1\rangle$  and thus the  dependence of $C_\infty$ with $\epsilon_0$ is asymmetric around this resonance, as has been already discussed in Ref.\onlinecite{gramajo_2018}. 
On the other hand, for the  ES resonance  ($\epsilon_{0}/\omega=3.75$), the entanglement generation for low $A$  is through the standard R3L mechanism with resonant pumping  $|0\rangle \leftrightarrow |2\rangle$  followed by the fast decay $|2\rangle \rightarrow |1\rangle$. In this case, the dependence of $C_\infty$ with $\epsilon_0$  shows a narrow symmetric peak around this resonance.

\section{Conclusions}
\label{s5}

We have presented an off-resonance three level (O3L) mechanism for steady-state entanglement generation in strongly driven qubits. The mechanism does not require  the fine tuning  of specific photon resonances as in the standard R3L mechanism, and it is not limited to a certain range of driving amplitudes, as in the N2L mechanism \cite{gramajo_2018}.
Unlike  these cases,  the O3L mechanism is efficient to generate  steady entanglement in a wide range of parameters, above a threshold of  bath coupling asymmetry and driving amplitude.

The proposed  strategy  for entanglement generation is based on the same principle   used for  microwave cooling of superconducting qubits in \cite{valenzuela2006}, which  essentially consists of the excitation to a higher level through a non-resonant process based on Landau-Zener transitions, plus the tuning of a fast relaxation channel to the desired final state.  In the microwave cooling case this protocol has been very efficient to pump population from the first excited state to the ground state, significantly lowering the effective temperature of the qubit. In our case, the O3L mechanism achieves the  active pumping of population from the separable ground state to an entangled excited state.
 
Small gap superconducting qubits are a good architecture to study quantum dynamics and quantum control processes based on Landau-Zener transitions \cite{oliver_2005,oliver_2009,valenzuela2006,campbell2020,gramajo_2020}. 
For instance,  Campbell {\it et al} \cite{campbell2020} have shown recently that through the use of non resonant Landau-Zener transitions is possible to achieve high-fidelity single qubit operations in these devices. Circuits of coupled small gap superconducting qubits, with added control of the system-bath coupling, are possible candidates for the implementation of the entanglement generation mechanism proposed here.

\section*{Acknowlegments}
We acknowledge support from CNEA, CONICET (PIP11220150100756), UNCuyo (P 06/C591) and ANPCyT (PICT2016-0791).

\appendix
\renewcommand\thefigure{\thesection.\arabic{figure}} 
\section{\label{sec:A} Floquet-Markov Master Equation}
\setcounter{figure}{0} 

As mentioned in Sec.\ref{s2} the open system dynamics can be described by the global Hamiltonian ${\cal H}(t)={H}_{s}(t) + {H}_{b} + {H}_{sb}$,
where ${H}_{s}(t)={H}_{0} + {V}(t)$ the Hamiltonian of two coupled qubits $H_0$ driving by a periodic external fields ${V}(t)$. Since ${H}_{s}(t)={H}_{s}(t+ \tau )$ is periodic in time, with  $\tau=2\pi/\omega_{0}$ the driving period, it is customary to employ the Floquet formalism
  to solve the dynamics. \cite{shirley_1965,grifoni_1998,grifoni_2010,ferron_2010} 
In the Floquet formalism the solutions of the   time dependent Schr\"odinger equation are of the form $|\Psi_\alpha(t)\rangle=e^{i\gamma_\alpha t/\hbar}|u_\alpha(t)\rangle$, where the  Floquet states $|u_\alpha(t)\rangle$ satisfy $|u_\alpha(t)\rangle$=$|u_\alpha(t+ \tau)\rangle = \sum_K |u_\alpha(K) \rangle e^{-iK\omega t}$ are eigenstates of 
$[{\cal H} (t)- i \hbar \partial/\partial t ] |u_\alpha(t)\rangle= \gamma_\alpha |u_\alpha(t)\rangle$, with $\gamma_\alpha$ the associated quasienergy.

We consider a bosonic thermal bath at temperature $T_{b}$ described by the usual harmonic oscillators Hamiltonian ${H}_{b}$, which is linearly coupled to  the two-qubits system in the form ${H}_{sb}=g\mathcal{A}\otimes\mathcal{B}$, with $g$ the coupling strength, $\mathcal{B}$ an observable of the bath and $\mathcal{A}$ an observable of the system as defined in Eq.\eqref{eq:sys_obs}.  In what follows we consider a bath with an ohmic spectral density $J(\Omega)=\gamma\Omega e^{-|\Omega|/\omega_{c}}$, with  $\omega_{c}$ the cutoff frequency.

The dynamics of the composed system is ruled by the von Neumann equation
  \begin{equation}
	\begin{aligned}
	\dot{\rho}_{tot}(t)=-\frac{i}{\hbar} \left[{\mathcal{H}}(t), \rho_{tot}(t)\right];
	\label{eq:A2}
	\end{aligned}
\end{equation} which after tracing over  the degrees of freedom of the bath becomes an equation for the  evolution of  the two-qubits  reduced density matrix $\rho(t)=Tr_{b}\left(  \rho_{tot} (t)\right)$,
  \begin{equation}
	\begin{aligned}
	\dot{\rho}(t)=-\frac{i}{\hbar} {\rm Tr}_{b} \left( \left[ {\mathcal{H}}(t), \rho_{tot}(t)\right] \right). 
	\label{eq:A3}
	\end{aligned}
\end{equation}  
After expanding $\rho(t)$ in terms of  the time-periodic  Floquet basis $\{|u_{\alpha}(t)\rangle\}$, 
(${\alpha}, {\beta}=0,1,2,3$)
\begin{equation}
\rho_{\alpha\beta}(t)=\langle u_{\alpha}(t)|\rho(t)|u_{\beta}(t)\rangle\;,
\end{equation}
 the Born (weak coupling) and Markov (local in time) approximations for the time evolution  are performed.
In this way, the Floquet-Markov Master equation   \cite{grifoni_1998,grifoni_2010,kohler_1997,kohler_1998,breuer_2000,ketzmerich_2009,ferron_2016,fazio_2013,solinas_2014} is 
obtained:
\begin{equation}
	\begin{aligned}
	\dot{\rho}_{\alpha\beta}(t)&=-i(\gamma_\alpha-\gamma_\beta)\rho_{\alpha\beta}-\sum_{\alpha'\beta'} \mathcal{L}_{\alpha\beta,\alpha'\beta'} (t)\rho(t)_{\alpha'\beta'},\\
	 \mathcal{L}_{\alpha\beta,\alpha'\beta'} (t)&=\sum_{Q}  \mathcal{L}^{Q}_{\alpha\beta,\alpha'\beta'} e^{-iQ\omega t},
	\label{eq:A4}
	\end{aligned}
\end{equation} with $ \mathcal{L}_{\alpha\beta,\alpha'\beta'} (t)$ the transition rates and $Q\in\mathbb{Z}$.  The Fourier coefficients are defined as
\begin{equation} 
	\begin{aligned}
        	\mathcal{L}^{Q}_{\alpha\beta,\alpha'\beta'} &=  \sum_{K} ( \delta_{\beta\beta'}\sum_{\eta} g^{K}_{\eta\alpha'} A^{K+Q}_{\alpha\eta}A^{K}_{\eta\alpha'}   \\
	& +  \delta_{\alpha\alpha'}\sum_{\eta} g^{-K}_{\eta\beta'} A^{K+Q}_{\eta\beta}A^{K}_{\beta'\eta}   \\
	& - \left(g^{K}_{\alpha\alpha'} + g^{-K-Q}_{\beta\beta'}\right) A^{K}_{\alpha\alpha'}A^{K+Q}_{\beta'\beta}),
	\label{eq:A5}
	\end{aligned}
\end{equation} with $g^{K}_{\alpha\beta} = J(\gamma_{\alpha\beta} + K\omega ) n_{th} (\gamma_{\alpha\beta} + K\omega)$, and $\gamma_{\alpha\beta} = \gamma_{\alpha} - \gamma_{\beta}$ and $K\in\mathbb{Z}$.
The thermal occupation is given by the  Bose-Einstein function 
 $n_{th}(x)=1/(e^{x/k_{B}T} - 1)$. Each $A^{K}_{\alpha\beta}$ is a transition matrix element in the Floquet basis, defined as $A^{K}_{\alpha\beta}=\sum_{L} \langle u_{\alpha} (L)|\mathcal{A}|u_{\beta}(L+K)\rangle$, with $|u_{\alpha}(L)\rangle$ the Fourier component of the Floquet state, $L\in\mathbb{Z}$.

By considering that the time scale $t_{r}$ for full relaxation is $t_{r} \gg \tau$, the transition rates  $ \mathcal{L}_{\alpha\beta,\alpha'\beta'} (t)$  can be thus  approximated by their average over one period $\tau$,  $ \mathcal{L}_{\alpha\beta,\alpha'\beta'} (t)\sim  {\mathcal{L}}^{Q=0}_{\alpha\beta,\alpha'\beta'}$ \cite{kohler_1997,kohler_1998}, obtaining
\begin{equation}
	\begin{aligned}
    \mathcal{L}^{Q=0}_{\alpha\beta,\alpha'\beta'} &=  \delta_{\beta\beta'} \sum_{\eta} {R}_{\eta\eta,\alpha'\alpha}    + \delta_{\alpha\alpha'}\sum_{\eta} ({R}_{\eta\eta,\beta'\beta})^{*}     \\
	&- {R}_{\alpha\beta,\alpha'\beta'} - ({R}_{\beta\alpha,\beta'\alpha'})^{*},
	\label{eq:A6}
	\end{aligned}
\end{equation} where the rates
\begin{equation}
	\begin{aligned}
	{R}_{\alpha\beta,\alpha'\beta'} &= \sum_{K} g_{\alpha\alpha'}^{K}  A^{K}_{\alpha\alpha'} \left( A^{K}_{\beta\beta'}\right)^{*},
	\label{eq:A7}
	\end{aligned}
\end{equation}  can be interpreted as sums of K-photon exchange terms.

The numerical procedure  is as  follows. First, the Floquet components  $|u_\alpha(K)\rangle$ are obtained by solving the unitary evolution and with
them,  the rates ${R}_{\alpha\beta\alpha'\beta'}$ and $\mathcal{L}^{Q=0}_{\alpha\beta,\alpha'\beta'}$ are computed. The
time dependent solution of $\rho_{\alpha\beta}(t)$ and the steady state $\rho_{\alpha\beta}(t\rightarrow\infty)$ are finally calculated as described in Ref.\cite{ferron_2016}.

\section{\label{sec:B} Eigenstates of $H_{0}$}

In this section, we give analytical expressions for  the eigenstates of the undriven-system $H_{0}$ using perturbation theory in the parameters $\Delta_i$. 

We start by writing the Hamiltonian  Eq.(\ref{h0}) as $H_0= H_0^ {(0)} + H_1$, being 
\begin{equation}
 {H}_{0}^{(0)} = \sum^{2}_{i=1}-\frac{\epsilon_{0}}{2}\sigma_{z}^{(i)}  -\frac{J}{2}\left(\sigma^{(1)}_{+}\sigma^{(2)}_{-} + \sigma^{(1)}_{-}\sigma^{(2)}_{+}\right),
 \label{h00}
 \end{equation}

and 
\begin{equation}
H_1 = \sum^{2}_{i=1} - \frac{\Delta_{i}}{2}\sigma_{x}^{(i)}.
\end{equation}

The eigenstates of $H_0^{(0)}$, spanned in the basis  of  $\sigma^{(1)}_{z}\otimes\sigma^{(2)}_{z}$, can be found by direct diagonalization, being 
the two Bell states $|e^{(0)}_{\pm}\rangle= \frac{1}{\sqrt{2}}(|01\rangle \pm|10\rangle)$ with eigenenergies $E^{(0)}_{e\pm}=\mp |J|/2$ and the two separable  eigenstates $|s^{(0)}_0\rangle=|00\rangle$ and  $|s^{(0)}_1\rangle=|11\rangle$, with eigenenergies $E^{(0)}_{s_0}=-\epsilon_0$ and $E^{(0)}_{s_1}=\epsilon_0$, respectively.
As we have mentioned in the main text,  for  both signs of $J$, the ground state of $H_0^{(0)}$ is entangled ($|e^{0}_{\mp}\rangle$)  for $|\epsilon_0|<|J|/2$ and separable ($|s^{0}_0\rangle$)  for $|\epsilon_0|>|J|/2$.

For  $\Delta_1,\Delta_2\ll|\epsilon_0|$, as we assumed  in the present analysis,  $H_1$ can be considered  as a  perturbative  term.
Straightforward calculations give to first order
in perturbation theory:
\begin{equation}
  \begin{aligned}
           &|s_{0}\rangle = |s^{(0)}_{0}\rangle+   \frac{\Delta_{-}}{ \epsilon_0 + J/2} |e^{(0)}_{-}\rangle  + \frac{ \Delta_{+}}{ \epsilon_0 -J/2} |e^{(0)}_{+}\rangle,  \\
           &|s_{1}\rangle = |s^{(0)}_{1}\rangle+   \frac{\Delta_{-}}{ \epsilon_0 - J/2} |e^{(0)}_{-}\rangle  - \frac{ \Delta_{+}}{ \epsilon_0 + J/2} |e^{(0)}_{+}\rangle, \\
           & |e_{-}\rangle=   |e^{(0)}_{-}\rangle - \frac{\Delta_{-}}{ \epsilon_0 + J/2} |s^{(0)}_{0}\rangle - \frac{\Delta_{-}}{ \epsilon_0 - J/2} |s^{(0)}_{1}\rangle  \\
           & |e_{+}\rangle=   |e^{(0)}_{+}\rangle - \frac{\Delta_{+}}{ \epsilon_0 - J/2} |s^{(0)}_{0}\rangle + \frac{\Delta_{+}}{ \epsilon_0 + J/2} |s^{(0)}_{1}\rangle  , 
           \end{aligned}
 \label{eq:eigho}
\end{equation}
where we define  $\Delta_{\pm}= \frac{\Delta_1 \pm \Delta_2}{ 2 \sqrt{2}}$.

\section{\label{sec:C} Transition rates}

With Eq.(\ref{eq:eigho}) at hand, one can compute  the transition rates $\Gamma_{f\leftarrow i} \equiv \Gamma_{fi}$ using the Fermi Golden rule (FGR),

\begin{equation}
\begin{aligned}
    \Gamma_{fi} = \frac{2\pi}{\hbar} g(E_{if})|\langle i |\mathcal{A} | f \rangle|^{2} ,
\end{aligned}
\label{eq:fermi_app}
\end{equation} where  the indexes $i$ and $f$ indicate the initial and final states, respectively,  $E_{if}=E_i-E_f$,  $\mathcal{A}$ is an observable of the system defined in Eq.\eqref{eq:sys_obs} and $g(E)= n_{th}(E)J(E/\hbar)$, is written as the product of the Bose-Einstein function  $n_{th}$  and $J(\Omega)$, the spectral density of the bath given in App.\ref{sec:A}.

Using Eq.\eqref{eq:fermi_app}, we obtain
\begin{equation}
\begin{aligned}
&\Gamma_{12}\sim \frac{2\pi}{\hbar} g(E_{21}) \gamma_1^{2}  \left[(1- \xi) + 2 (1 + \xi) \frac{ \Delta_{-} \Delta_{+}}{({\epsilon_0}^2 -  \frac{J^2}{4})} \right]^2,\\
&\Gamma_{01} \sim \frac{2\pi}{\hbar} g(E_{10}) \gamma_1^{2} \left[(1- \xi) \frac{\Delta_{+}}{(\epsilon_0 - \frac{J}{2})} + (1 + \xi) \frac{ \Delta_{-}}{(\epsilon_0 + \frac{J}{2})} \right]^2  ,\\
&\Gamma_{13} \sim \frac{2\pi}{\hbar} g(E_{31}) \gamma_1^{2} \left[ (1- \xi) \frac{\Delta_{+}}{(\epsilon_0 + \frac{J}{2})} + (1 + \xi) \frac{ \Delta_{-}}{(\epsilon_0 -\frac{J}{2})} \right]^2,\\
&\Gamma_{02} \sim \frac{2\pi}{\hbar} g(E_{20}) \gamma_1^{2} \left[ (1+ \xi) \frac{\Delta_{+}}{(\epsilon_0 - \frac{J}{2})} + (1 - \xi) \frac{ \Delta_{-}}{(\epsilon_0 +  \frac{J}{2})} \right]^2,\\
& \Gamma_{23} \sim \frac{2\pi}{\hbar} g(E_{32}) \gamma_1^{2} \left[ (1+ \xi) \frac{\Delta_{+}}{(\epsilon_0 + \frac{J}{2})} + (1 - \xi) \frac{ \Delta_{-}}{(\epsilon_0 - \frac{J}{2})} \right]^2, 
\end{aligned}
\label{eq:rates_app}
\end{equation} where the labeling corresponds to the eigenstates ordering given for $J<0$, while  $1\leftrightarrow 2$ for $J>0$, as in this case  the first and second excited states are   $|1\rangle= |e_{+}\rangle$ and $|2\rangle=|e_{-}\rangle$, respectively. 
These expressions correspond to transition rates between the eigenstates of $H_0$. Thus, they are  adequate for describing the relaxation dynamics in the undriven system or for weak driving amplitudes.

In general, the relaxation dynamics of the strongly driven system should be discussed  in terms of the transition rates $\Gamma_{\alpha\beta}$ between Floquet states, which can be obtained as follows.
After performing  the secular approximation \cite{gramajo_2018}, Eq.\eqref{eq:A4} transforms into a Lindblad-type equation, given by
\begin{equation}
	\begin{aligned}
	\dot{\rho}&= -i [ H_{s}(t),\rho] + \sum_{\alpha\beta}  \Gamma_{\alpha\beta} \left( L_{\alpha\beta}\rho L^{\dag}_{\alpha\beta} - \frac{1}{2} \{ L^{\dag}_{\alpha\beta}L_{\alpha\beta},\rho \}\right),
	\label{eq:C1}
	\end{aligned}
\end{equation} where $L_{\alpha\beta} = |u_{\alpha}(t)\rangle \langle u_{\beta}(t)|$ are the corresponding jump operators, and the transition rates $\Gamma_{\alpha\beta}=R_{\alpha\alpha,\beta\beta}$ can be written as 
\begin{equation}
	\begin{aligned}
	\Gamma_{\alpha\beta} &= \sum_{n}  \Gamma_{\alpha\beta}^{(n)},\\
	\Gamma_{\alpha\beta}^{(n)}&=g(\gamma_{\alpha\beta}+n\omega)|A^{n}_{\alpha\beta}|^{2},
	\label{eq:C2}
	\end{aligned}
\end{equation} being   $\gamma_{\alpha\beta} = \gamma_{\alpha} - \gamma_{\beta}$ the quasienergies difference.

\begin{figure}[!htb]
\centering
\includegraphics[width=7.5cm]{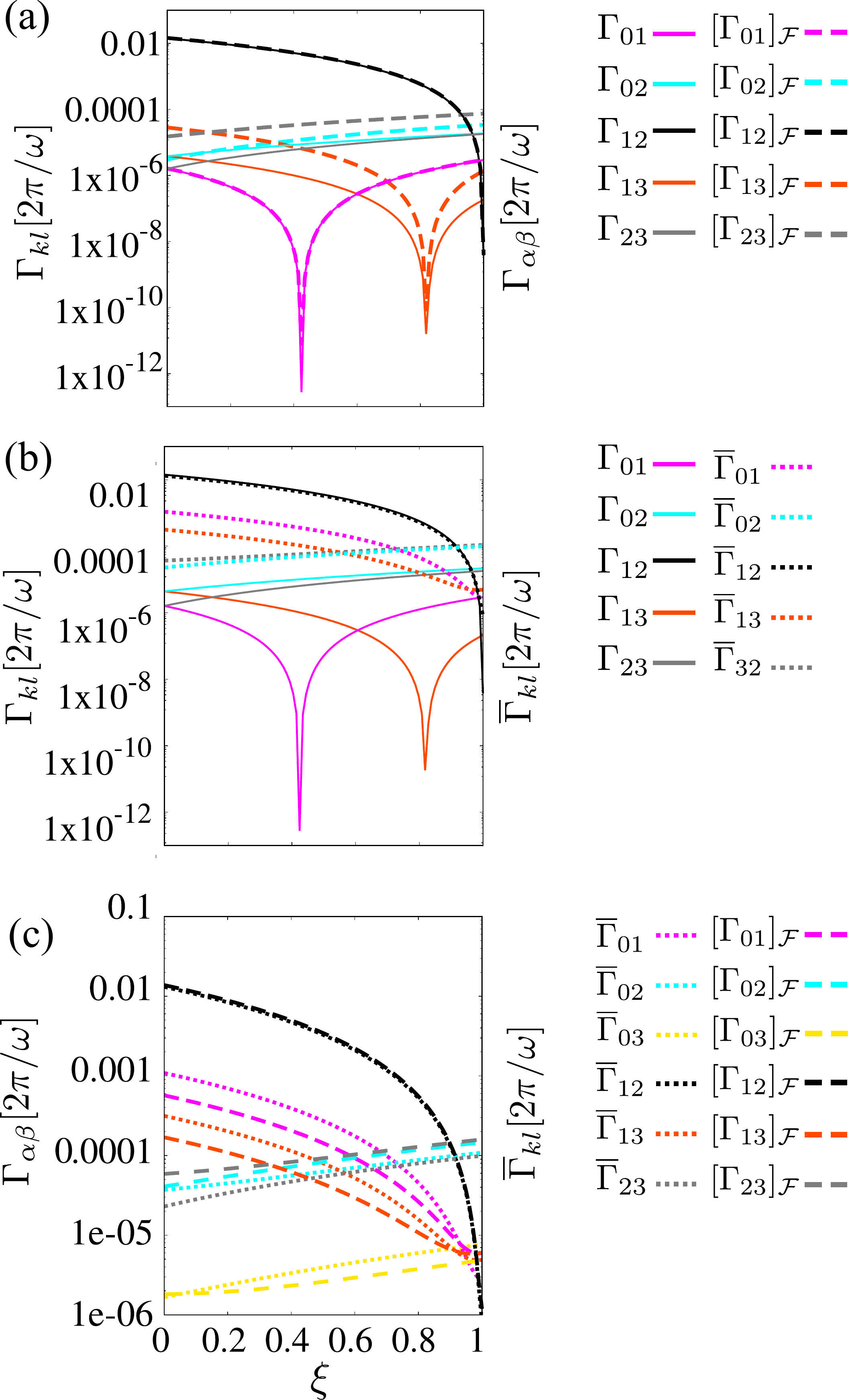}
\caption{Plot of the transition rates $\Gamma_{kl}$, $\overline{\Gamma}_{kl}$ and $\Gamma_{\alpha\beta}$ as a function of $\xi$ for the off-resonance condition $\epsilon_{0}/\omega=3.7$. (a) The rates $\Gamma_{kl}$ are plotted in the eigenstates basis using the analytical estimates given by Eq.\eqref{eq:rates_app} (bold lines), while  $\Gamma_{\alpha\beta}$ are numerically computed in the Floquet basis (dashed lines). Both results correspond to $A/\omega=0$. (b)  Rates $\Gamma_{kl}$ plotted in (a) and rates $\overline{\Gamma}_{kl}$ computed using Eq.\eqref{eq:C7} (dotted lines) for  $A/\omega=3.8$. (c) Rates $\overline{\Gamma}_{kl}$ plotted in (b) and rates  ${\Gamma}_{\alpha\beta}$ computed numerically, both for $A/\omega=3.8$. Other parameters are the same as in Fig.\ref{fig:2}.}
\label{fig:C1}
\end{figure}

When the driving is weak, for $A\rightarrow 0$, the Floquet states tend to the eigenstates of $H_0$, $|u_{\alpha}(t)\rangle \rightarrow |i\rangle$. Thus, for $A=0$ the Floquet rates $\Gamma_{\alpha\beta}^{(A=0)}$ coincide with the eigenstate rates $\Gamma_{ij}$.
In Fig.\ref{fig:C1}(a) we compare the approximate expressions obtained from Eq.\eqref{eq:rates_app} for the rates $\Gamma_{ij}$ with
the  Floquet rates $\Gamma_{\alpha\beta}$ computed  numerically  for $A=0$ and $\epsilon_{0}/\omega=3.7$. Since $\Delta_1,\Delta_2\ll|\epsilon_0|$, there is a good agreement between the rates approximated by Eq.\eqref{eq:rates_app} and the exact numerical rates.

In the following we  derive effective expressions for the rates $\Gamma_{kl}$ between the eigenstates for finite driving amplitudes $A\not=0$ in terms of the Floquet transition rates   $\Gamma_{\alpha\beta}$. We start by rewriting Eq.\eqref{eq:C1} in the interaction picture, i.e. $\tilde{\rho} = U^{\dag}(t) \rho U(t)$ with  $U(t)= \sum_{\alpha} e^{-i\gamma_{\alpha}t} |u_{\alpha}(t)\rangle\langle u_{\alpha}(0)|$, 
\begin{equation}
	\begin{aligned}
	\dot{\tilde{\rho}}&= \sum_{\alpha\beta}  {\Gamma}_{\alpha\beta} \left( \tilde{L}_{\alpha\beta}\tilde{\rho} \tilde{L}^{\dag}_{\alpha\beta} - \frac{1}{2} \{ \tilde{L}^{\dag}_{\alpha\beta}\tilde{L}_{\alpha\beta},\tilde{\rho} \}\right),
	\label{eq:C3}
	\end{aligned}
\end{equation} with $\tilde{L}_{\alpha\beta}= |u_{\alpha}(0)\rangle \langle u_{\beta}(0)|$.

After performing the basis change (with $H_{0}|i\rangle = E_{i} |i\rangle$)
\begin{equation}
	\begin{aligned}
	\tilde{L}_{\alpha\beta}&= \sum_{ij} |i\rangle\langle i| u_{\alpha}(0)\rangle \langle u_{\beta}(0)|j\rangle\langle j| ,\\
	&= \sum_{ij} \langle i| u_{\alpha}(0)\rangle \langle u_{\beta}(0)|j\rangle L_{ij},
	\label{eq:C4}
	\end{aligned}
\end{equation} with ${L}_{ij}= |i \rangle \langle j|$, and replacing into Eq.\eqref{eq:C3}, we get
\begin{equation}
	\begin{aligned}
	\dot{\tilde{\rho}}&= \sum_{ijkl}  \overline{\Gamma}_{ij,kl} \left(  {L}_{ij}\tilde{\rho} {L}^{\dag}_{kl} - \frac{1}{2} \{ {L}^{\dag}_{kl}{L}_{ij},\tilde{\rho} \}\right),
	\label{eq:C5}
	\end{aligned}
\end{equation} with
\begin{equation}
	\begin{aligned}
	\overline{\Gamma}_{ij,kl} = \sum_{\alpha\beta} \langle i | u_{\alpha}(0)\rangle \langle u_{\beta}(0)| j \rangle \langle l | u_{\beta}(0)\rangle \langle u_{\alpha}(0)| k \rangle \Gamma_{\alpha\beta}.
	\label{eq:C6}
	\end{aligned}
\end{equation} Further assuming that only diagonal terms dominate when the system is fully relaxed (as each Floquet state has principal weight
on the associated eigenstate) the relevant rates in Eq.\eqref{eq:C6} are $\overline{\Gamma}_{ij,ij} \equiv \overline{\Gamma}_{ij}$, which read 
\begin{equation}
	\begin{aligned}
	\overline{\Gamma}_{ij} = \sum_{\alpha\beta} |\langle i | u_{\alpha}(0)\rangle|^{2} |\langle u_{\beta}(0)| j \rangle|^{2}  \Gamma_{\alpha\beta}.
	\label{eq:C7}
	\end{aligned}
\end{equation}

The obtained  $\overline{\Gamma}_{ij}$ allow to define effective transition rates between eigenstates for finite driving amplitudes $A$. In principle,  the rates $\overline\Gamma_{kl}$ given by Eq.\eqref{eq:C7} can be different than  the rates  for $A=0$ given  by Eq.\eqref{eq:fermi_app}.
This is illustrated in Fig.\ref{fig:C1}(b) where we plot the FGR estimates, Eq.\eqref{eq:rates_app},
and the rates  $\overline\Gamma_{kl}$ computed numerically for $A/\omega=3.8$. In the comparison of $A=0$ rates with the $A\not=0$ effective rates we find that
in spite of   evident differences, the  most relevant    transition rates: $\Gamma_{12}$, that dominates the dynamics for $\xi<\xi_c$, and  $\Gamma_{02}$, $\Gamma_{23}$, that dominate the dynamics for $\xi > \xi_c$, have a similar dependence with $\xi$. One  thus can conclude that the strength of
this dominant decay channels is mainly  determined by the degree of asymmetry in the qubits-bath couplings, $1-\xi$, in agreement with the estimates given  in Eq.(\ref{eq:rates_app}).

Finally  in Fig.\ref{fig:C1}(c), we compare  the effective rates $\overline\Gamma_{kl}$ given by Eq.\eqref{eq:C7} with the Floquet
rates $\Gamma_{\alpha\beta}$, both  for $A/\omega = 3.8$,  $\epsilon_{0}/\omega = 3.7$ and $J/\omega = -2.5$.
We find that  most of the  rates labelled by the same indexes   follow a similar  trend with $\xi$ and are in   good  qualitative agreement. In particular, the  rates $\Gamma_{12}$, $\Gamma_{02}$ and $\Gamma_{23}$ which
 dominate the main relaxation processes associated to the O3L mechanisms studied in Sec.(\ref{s3}), are in  good  quantitative
 agreement with their counterparts computed  in the Floquet basis, $[\Gamma_{12}]_{\mathcal F}$, $[\Gamma_{02}]_{\mathcal F}$ and $[\Gamma_{23}]{\mathcal F}$, respectively.
 The similarity among $\overline{\Gamma}_{ij}$ and $[\Gamma_{\alpha\beta}]_{\mathcal F}$, reflect the fact that out of resonance the eigenstates and the Floquet states strongly overlap.

\section{\label{sec:D} Dependence on   the qubit-qubit coupling} 

\begin{figure}[!htb]
\centering
\includegraphics[width=5.5cm]{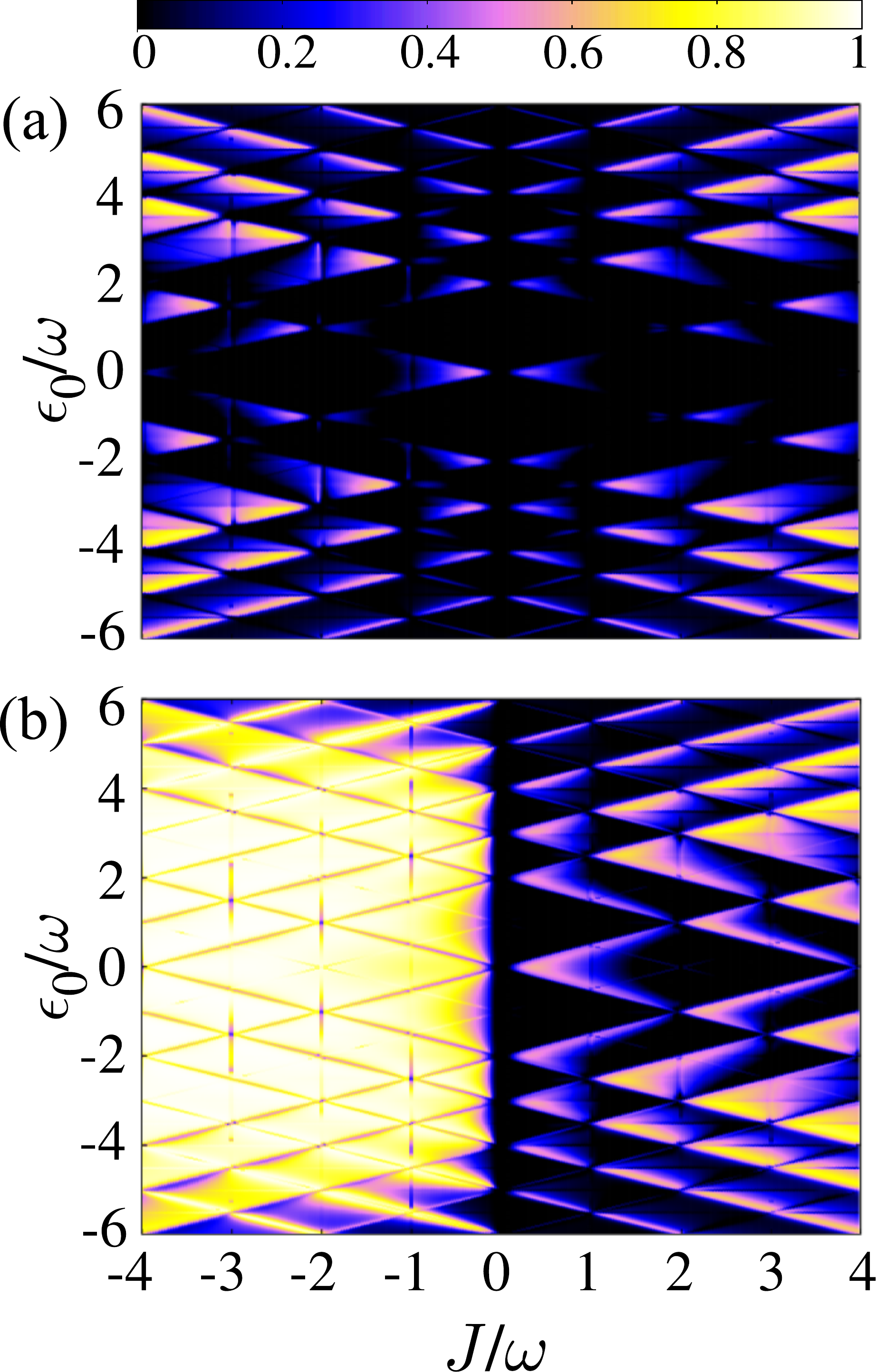}
\caption{Intensity plot of $C_{\infty}$ versus $J/\omega$ and $\epsilon_{0}/\omega$ for $\xi=1$ (a)  and  $\xi=0.1$ (b). In  both cases is $A/\omega=3.8$ while other parameters have  the same  values as in Fig.\ref{fig:2}.}
\label{fig:D2}
\end{figure}

A natural question that arises is how the  entanglement generation described in Sec.\ref{s2} is  modified as  the qubit-qubit interaction strength  $J$ changes. Figure \ref{fig:D2} shows the intensity plot of  $C_{\infty}$ as a function of  $J/\omega$ and $\epsilon_{0}/\omega$ for $\xi=1$ (a) and  $\xi=0.1$ (b). 
While  $C_{\infty}$ for  $\xi=1$ seems to be independent on  the sign of $J/\omega$, i.e.  $C_{\infty}(J) \sim C_{\infty} (-J) $,
for $\xi=0.1$ (Fig.\ref{fig:D2}(b)),  $C_{\infty}$ exhibits striking differences. Nevertheless, in the two cases we notice the relevance of the resonances among a separable and an entangled state, which correspond to the straight lines defined by $\epsilon_0 \pm |J|/2 = n\omega$ in Fig.\ref{fig:D2}. 

The behaviour shown  in Fig.\ref{fig:D2}(a) has been extensively discussed  in Ref.\onlinecite{gramajo_2018} and we refer the reader to this paper  for specific details.
To sumarize, the triangular-like structure of $C_{\infty}$ that shows enhanced  entanglement  at one side of the resonances defined by  $\epsilon_{0} - |J|/2 \sim m \omega$ (together with its symmetry  with the sign of $J/\omega$ and anti symmetry with $\epsilon_{0}/\omega$, respectively)  is related  to the aforementioned   N2L mechanism, mediated by the bath and the external driving. 
Notice that in this case there is not entanglement generated by the O3L mechanism. This is due to the fact that for $\xi=1$,  
the relaxation rates  that could contribute to populate in the steady state the entangled state $|1 \rangle \equiv | e_{-} \rangle$  for $J <0$ ($|1\rangle \equiv | e_{+} \rangle$
for $J >0$, see Fig.\ref{fig:1} (b)), satisfy  $\Gamma_{12}\sim 0$ for  both signs of $J$ (the analytical estimates for the rates  computed in the eigenbasis of $H_0$ and  for $J >0$   are given  by   Eq.(\ref{eq:rates})
but with the  sub-indexes  $ 1 \leftrightarrow 2$ interchanged, as  for $J >0$ the first  and second excited  states are $|1\rangle \equiv| e_{+} \rangle$ and  $|2\rangle \equiv| e_{-} \rangle$, respectively).

\begin{figure}[!htb]
\centering
\includegraphics[width=6.4cm]{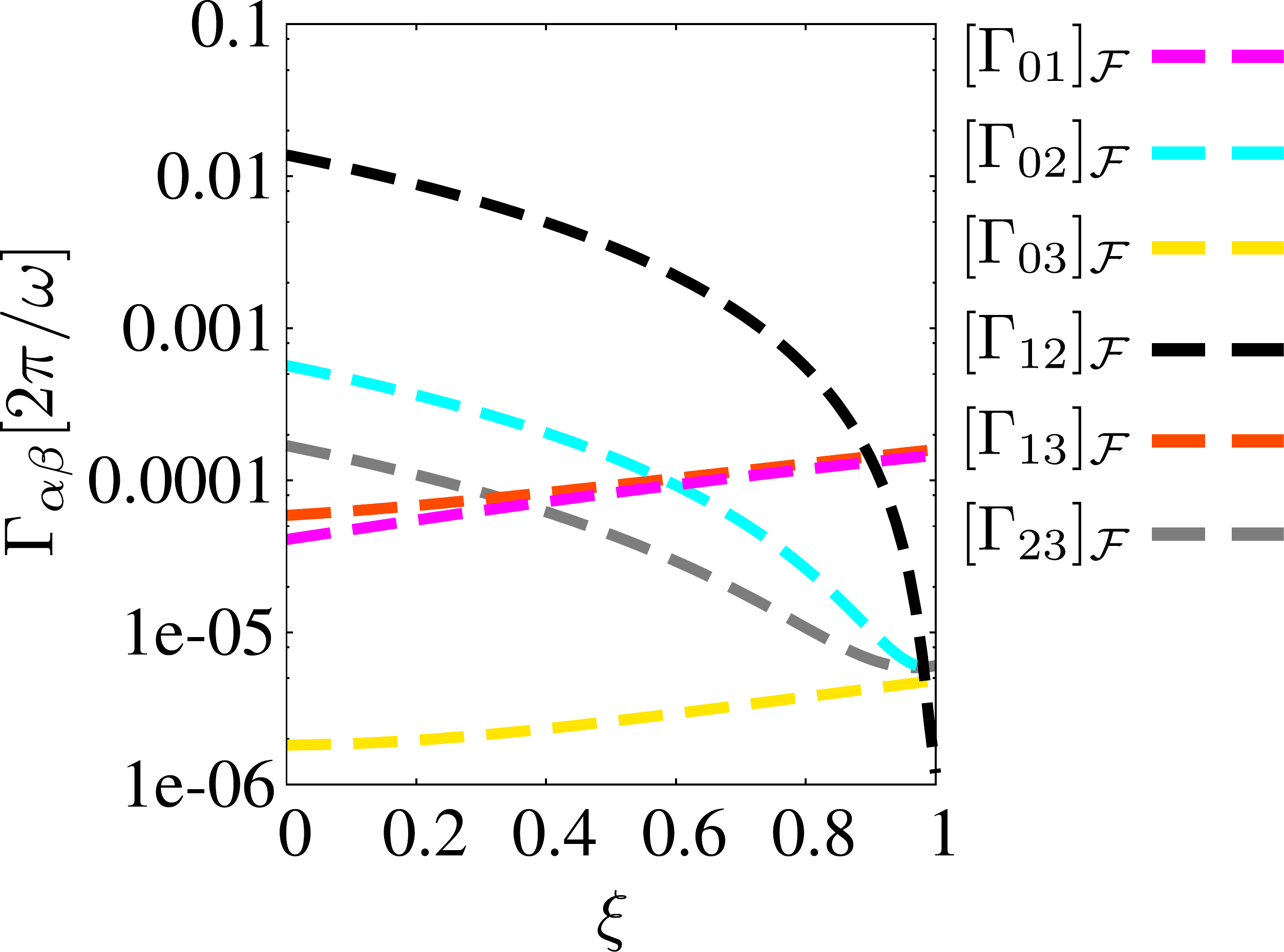}
\caption{Transition rates $\Gamma_{\alpha\beta}$, computed in the Floquet basis, as a function of $\xi$ for the  off-resonance case $\epsilon_{0}=3.7\omega$.  The results correspond to  $J/\omega= 2.5$ and $A/\omega=3.8$. Other parameters are the same as in Fig.\ref{fig:2}.}
\label{fig:D3}
\end{figure}

Indeed, for $J >0$ and $\xi=1$, the relevant relaxation rates  tend to populate the $|0\rangle \equiv| s_{0} \rangle$ separable state, in analogy with  the  case for  $J <0$ and $\xi=1$, previously discussed. 

Notice that as we have already argued and following   the analysis given in Appendix \ref{sec:C},
the  rate computed in the Floquet basis satisfies $[\Gamma_{12}]_{\mathcal{F}}\sim 0$  for $\xi=1$ and for  both signs of $J$, obeying
the same trend than the rate computed in the eigenbasis of $H_0$,  as Figs.\ref{fig:5} and \ref{fig:D3} display.

In the following we shall discuss the behavior observed in Fig.\ref{fig:D2}(b) for $\xi=0.1$.
The results for the negative branch  $J<0$ can be understood following the same reasoning  than in  the previous Sec.\ref{s2}, where we focused on  $J/\omega=-2.5$.
 The  value $C_{\infty} \lesssim 1$ is attained  for almost all the explored range
 of parameter space with   largest  value  $C_{\infty} \simeq 1$ corresponding to the conditions for  the O3L mechanism  previously described and  with  the  relaxation  rates  essentially  given  by  Eq.(\ref{eq:rates}).
 As in  the present case is $A= 3.8 \omega$, the entanglement creation by the O3L mechanism  fades out  outside  the wedge-shaped region delimited by the  lines  $J/ \omega = \pm (\epsilon_{0}/\omega - 3.8) $, as  outside this region   is  $ A < A_c$, and thus 
the  critical amplitude  $ A_c$ necessary to activate the O3L mechanism is not attained.

A visible   feature in Fig.\ref{fig:D2}(a) is that, superimposed on  the quasi-homogeneous pattern of $C_{\infty} \sim 1 $, straight lines  along
which is  $C_{\infty} \sim 0.5$ are clearly observed nearby to  the resonance conditions $\epsilon_{0} \pm |J|/2 \sim m \omega$, which have been discussed in Sec.\ref{s4}.

For the positive branch $J>0$ the  O3L mechanism is completely suppressed, and instead 
there is  a triangular-like structure  rather similar to Fig.\ref{fig:D2}(a), i.e. corresponding to entanglement generated by the N2L mechanism. This result can be easily explained as follows. As mentioned before, for $J > 0$, the first and second excited states are switched between each other $|e_{-}\rangle \leftrightarrow |e_{+}\rangle $ ($|1\rangle \leftrightarrow |2\rangle$).  As a consequence, for  $J >0$ and $\xi \rightarrow 0$, in addition to $[\Gamma_{12}]_{\mathcal{F}}$, the second 
relevant relaxation rate is $[\Gamma_{02}]_{\mathcal{F}}$.
Thereby, there are now two  relaxation mechanisms from the $|2\rangle$ state affecting the dynamics, the $|2\rangle \rightarrow |1\rangle$ transition and the $|2\rangle \rightarrow |0\rangle$ transition. This later mechanism tends to populate the ground state at long times.

\bibliography{references1}

\end{document}